\documentclass{jetpl}
\usepackage[koi8-r]{inputenc}
\usepackage{isotope}

\twocolumn

\lat

\title{Pauli-principle driven correlations in four-neutron nuclear decays}
\rtitle{Pauli-principle driven correlations in four-neutron nuclear decays}
\sodtitle{Pauli-principle driven correlations in four-neutron nuclear decays}

\author{
  P.\,G.\,Sharov\textsuperscript{*}\/\thanks{e-mail: sharovpavel@jinr.ru},
  L.\,V.\,Grigorenko\textsuperscript{*\dag\ddag},
  A.\,N.\,Ismailova\textsuperscript{*},
  M.\,V.\,Zhukov\textsuperscript{\S}
}

\rauthor{
  P.\,G.\,Sharov,
  L.\,V.\,Grigorenko,
  A.\,N.\,Ismailova,
  M.\,V.\,Zhukov
}

\sodauthor{
  Sharov,
  Grigorenko,
  Ismailova,
  Zhukov
}

\address{
  \textsuperscript{*}Flerov Laboratory of Nuclear Reactions, JINR,
  141980 Dubna, Russia\\
  \textsuperscript{\dag}National Research Nuclear University ``MEPhI'',
  115409 Moscow, Russia\\
  \textsuperscript{\ddag}National Research Centre ``Kurchatov Institute'',
  Kurchatov sq.~1,123182 Moscow, Russia\\
  \textsuperscript{\S}Department of Physics, Chalmers University of Technology,
  S-41296 G\"{o}teborg, Sweden
}

\dates{28th March 2019}{*}

\abstract{
  Mechanism of simultaneous non-sequential four-neutron ($4n$) emission
  (or ``true'' $4n$-decay) has been considered in phenomenological
  five-body approach.
  This approach is analogous to the model of the direct decay to the continuum often
  applied to $2n$- and $2p$-decays.
  It is demonstrated that $4n$-decay fragments should have specific energy and
  angular correlations reflecting strong spatial correlations of
  ``valence'' nucleons orbiting in their $4n$-precursors.
  Due to the Pauli exclusion principle, the valence neutrons are pushed to
  the symmetry-allowed configurations in the $4n$-precursor structure,
  which causes a ``Pauli focusing'' effect.
  Prospects of the observation of the Pauli focusing have been considered for the
  $4n$-precursors \isotope[7]{H} and \isotope[28]{O}.
  Fingerprints of their nuclear structure or/and decay dynamics are predicted.
}

\PACS{27.20.+n, 21.65.Cd, 21.45.-v, 25.70.Pq, 23.20.En}

\begin{document}

\maketitle

%===============================================================================

\textbf{Introduction.}
In the last decade there was a great progress in the studies of
three-body decays (e.g.\ two-proton radioactivity)~\cite{Pfutzner:2012}.
In contrast to ``conventional'' two-body decays, three-body decays encrypt
a lot of additional information in the momentum
(energy and angular) correlations of the decay products.
Theoretical studies indicate that both effects of the initial nuclear structure
and the decay mechanism may show up in the core+$n$+$n$ and core+$p$+$p$
fragment correlation patterns in
various ways~\cite{Grigorenko:2003b,Mukha:2006,Grigorenko:2007,Grigorenko:2007a,Mukha:2008,
  Grigorenko:2009,Grigorenko:2009c,Egorova:2012,Grigorenko:2013,Brown:2014,
  Brown:2015,Grigorenko:2018}.

With the development of experimental techniques, more and more ``complicated''
nuclear systems become available for studies.
One of such complicated cases are isotonic neighbors of the $4n$-halo systems
located beyond the neutron dripline, which are expected to have
narrow resonance ground state decaying via $4n$-emission.
The examples of such systems, which are now actively studied by experiment, are
\isotope[7]{H} and \isotope[28]{O}.
The $4n$-emission phenomenon is known to be widespread beyond the neutron
dripline, and other possible candidates for such a decay mode,
e.g.\ \isotope[18]{Be} can be mentioned.
Their ground states are expected to be unbound with $E_T \lesssim 2$ MeV
($E_T$ is energy above the $4n$ decay threshold), and the decay mechanism
can be assumed as ``true'' $4n$ emission:
there are no sequential neutron emissions,
which mean that all neutrons are emitted simultaneously.
For low decay energies with $E_T \lesssim 300$ keV such a decay mechanism
may lead to very long lifetimes characterized in terms of
$4n$ radioactivity~\cite{Grigorenko:2011}.
From the experimental side there are no studied examples of \(4n\) decay;
only \(4p\) emission from \isotope[8]{C}
has been studied so far~\cite{Charity:2011}.
However, in the latter case the decay has a mechanism of sequential
$2p$ emissions \(\isotope[8]{C}\to 2p+ (\isotope[6]{Be}\to 2p+\alpha)\),
which can be treated with conventional three-body methods.

In the $4n$-emission (core+$4n$ decay)
the five-body correlations encrypt
enormously more information compared to the three-body decay.
In five-body case
the complete correlation pattern is described by 8-dimensional space
compared to the 2-dimensional space in the three-body decay.
The core+$4n$ system permutation symmetries
should decrease the effective dimension
of the correlation space, but there should be still a lot.
The question can be asked here
``How we should look for physically meaningful signals in this wealth of
information?''

Few words about important techniques used in correlation phenomenology.
The most easy-to-understand way to access information provided by correlations
is the idea of Migdal-Watson (MW) approximation:
two-body excitation spectra can be studied by correlating two selected products
of few-particle decays \cite{Migdal:1955,Watson:1952}.
The method stemming from MW is application of intensity
(Hanbury-Brown-Twiss or HBT-like) interferometry to problems of
femtoscopy \cite{Lisa:2005}.
In certain cases effect of the third particle should be considered in HBT and
correlations become implicitly three-body \cite[and Refs.\ therein]{Shoppa:2000}.
In particle physics for three-body processes the Dalitz plot techniques provide
powerful means for studies of resonance properties in two-body subsystems and
spin-parity identification \cite{Tanabashi:2018}.
The idea of two-dimensional Dalitz plot is based on the fact that three-body
correlations in the decays are fully described by just two degrees of freedom.
These techniques are also often applied to three-body decays of nuclear systems,
e.g.\ \cite{Fynbo:2009,Revel:2018}.
In few-body physics it is popular to describe three-body correlation patterns
in terms of the energy-angular $\{\varepsilon, \theta_k \}$ correlations for
Jacobi momenta $\{\mathbf{p}_{x}, \mathbf{p}_{y} \}$ \cite{Pfutzner:2012}
\begin{align}
  \varepsilon &=  E_x/E_T \, ,\qquad \cos(\theta_k)=(\hat{\mathbf{p}}_{x},
                \hat{\mathbf{p}}_{y}) \label{eq:jac} \\
  E_T\! & =  E_x+E_y = p^2_x/2 \mu_x + p^2_y/ 2 \mu_y ,  \nonumber \\
  \mathbf{p}_{x} &= \mu_{ik} \left({\mathbf{p}_i/M_i} - {\mathbf{p}_k
                / M_k} \right) , \; \; \mu_{ik} = M_i M_k /(M_i + M_k),
                \nonumber \\
  \mathbf{p}_{y} &= \mu_{ikn}\! \left( {\mathbf{p}_i+\mathbf{p}_k \over M_i + M_k } -
                {\mathbf{p}_n \over M_n} \right)\!\! , \;
                \mu_{ikn}\! = \frac{ (M_{i}+M_{k}) M_n}{ M_i + M_k + M_n },\nonumber
\end{align}
for some selection of the Jacobi system defined by enumeration of three
particles $i \neq k \neq n$ with masses $M_i$ and momenta vectors
$\mathbf{p}_i$. More complicated forms of three-body correlations become
available in the studies of direct reactions producing three and more particles
in the final state \cite{Chudoba:2018}.

In this work we would like to extend these practices to five-body core+$4n$
decays.
It is demonstrated that core-$n$ and $n$-$n$ energy and angular
distributions may provide important information on the dominant configurations
involved in the \(4n\) decay.
It is found, however, that \emph{isolated} energy
and angular distributions are not very expressed and can be affected in
complicated way by different aspects of structure and dynamics.
In contrast, the studies of \emph{correlated} $\{i$-$k,n$-$m\}$ patterns of
energy and angular distributions may give unique signatures for
orbital configurations involved in the decay
(here numbers $i \neq k$, $n\neq m$
enumerate different selections of pairs of the decay products).

%===============================================================================

\begin{figure}[!t]
\centering
\includegraphics[width=0.23\textwidth]{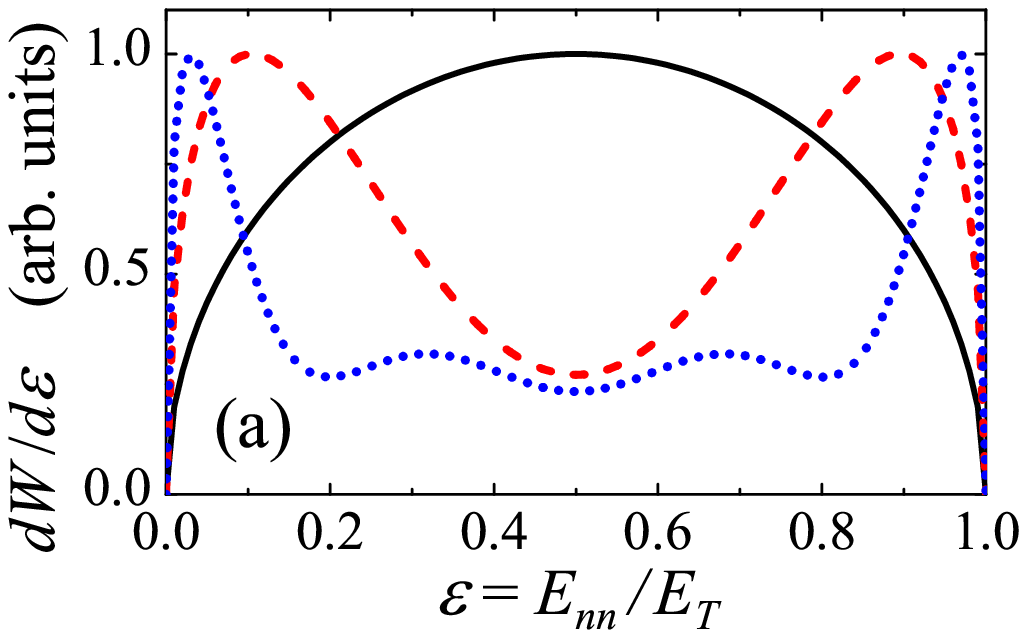}%
\includegraphics[width=0.23\textwidth]{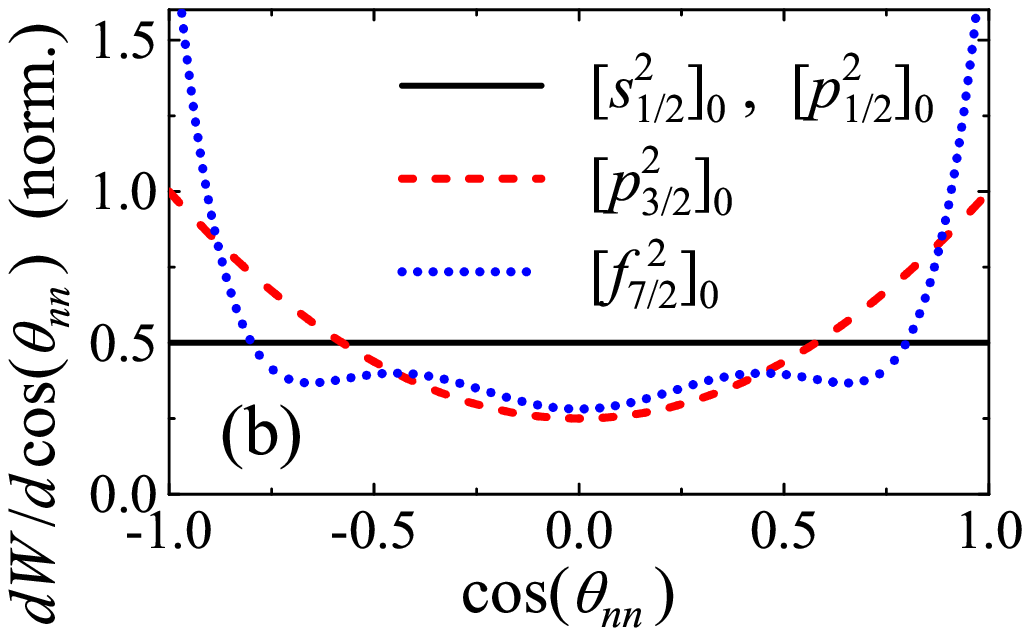}
\includegraphics[width=0.23\textwidth]{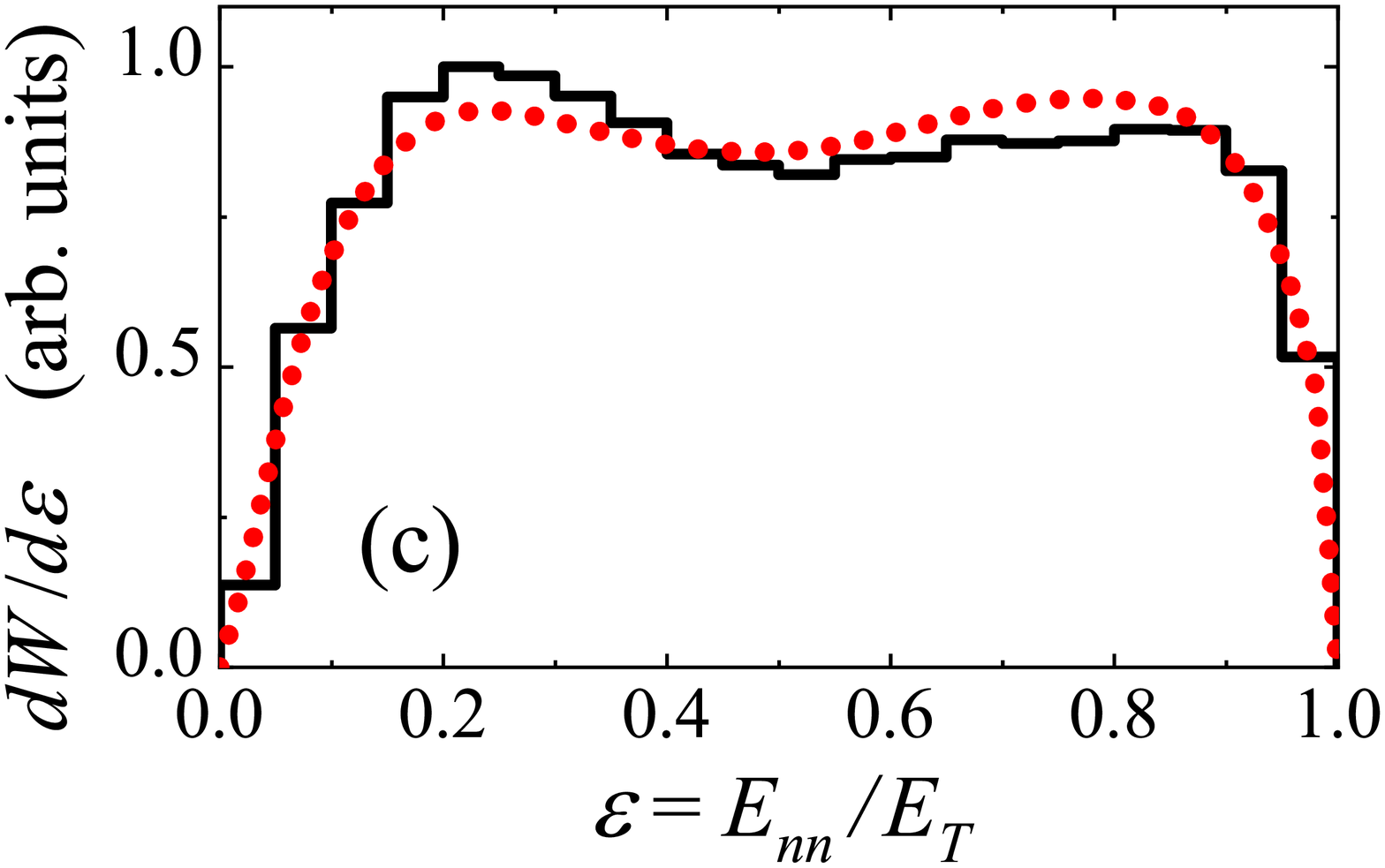}%
\includegraphics[width=0.23\textwidth]{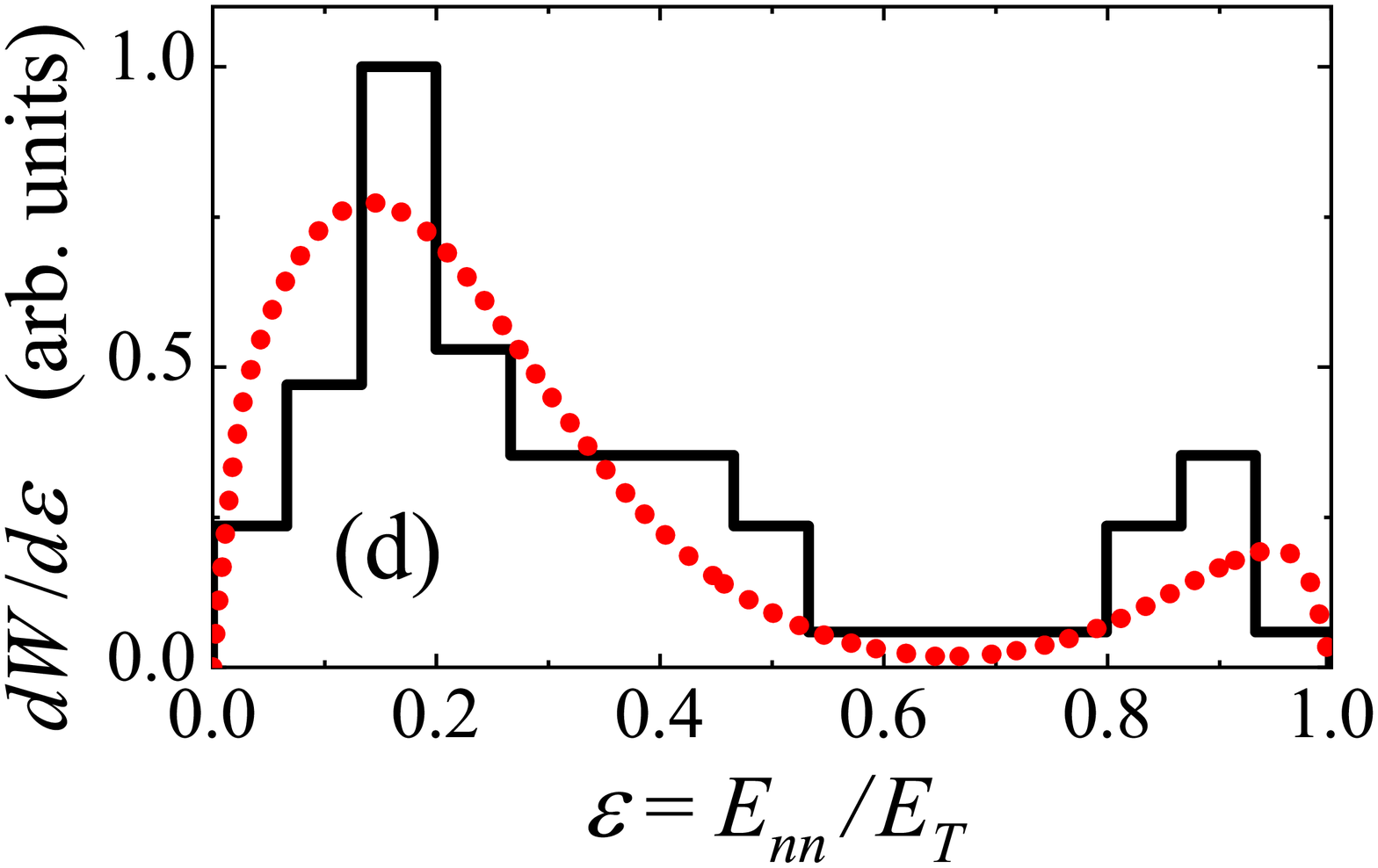}
  \caption{\label{fig:pauli-mom-3} Fig.~\ref{fig:pauli-mom-3}.
    Pauli focusing for three-body decays with valence $2n$ in configurations
    ${[s^2_{1/2}]}_0$, ${[p^2_{1/2}]}_0$, ${[p^2_{3/2}]}_0$, and ${[f^2_{7/2}]}_0$.
    Panel (a) shows $n$-$n$  energy distribution,
    panel (b) shows $n$-$n$ angular distribution obtained neglecting
    $n$-$n$ final state interaction.
    Panels (c) and (d) show the energy $p$-$p$ distributions observed for
    \isotope[6]{Be}~\cite{Egorova:2012} and \isotope[45]{Fe}~\cite{Grigorenko:2009}.
    Red dotted curves in panels (c,d) correspond to complete three-body
    theoretical calculation presented in
    papers~\cite{Grigorenko:2003b,Grigorenko:2009c}.}
\end{figure}
%-------------------------------------------------------------------------------

\textbf{The concept of ``Pauli focusing''.}
This concept was proposed in \cite{Danilin:1988} and further discussed e.g.\
in \cite{Zhukov:1993,Mei:2012} for the bound state structure of
three-body core+$n$+$n$ systems.
It was demonstrated that due to the Pauli exclusion principle,
the population of orbital configurations $[l_{j_1} \otimes l_{j_2}]_J$ for
the valence nucleons  may induce strong spatial correlations depending on
the specific values of $j_1$, $j_2$, and $J$.
Various forms of such correlations were actively discussed
as an integral part of the two-nucleon halo phenomenon
and now they seem to be well experimentally confirmed.

For bound states the theoretically predicted Pauli focusing correlations are
``hidden'' in the nuclear interior and can be accessed experimentally only in
some indirect way.
In contrast, in the three-body decay process these internal correlations may
directly exhibit themselves in the momentum distributions of the decay products.
This is illustrated in Fig.~\ref{fig:pauli-mom-3} for the three-body decays of
the $[l^2_j]_J$ orbital configurations (the direct decay into continuum is assumed).
The $[s^2_{1/2}]_0$ and $[{p^2}_{1/2}]_0$ configurations produce phase-space
function $\sqrt{\varepsilon(1-\varepsilon)}$ in the nucleon-nucleon channel and
isotropic angular distribution between nucleons, while such configurations as
$[p^2_{3/2}]_0$ and $[f^2_{7/2}]_0$ induce strong $2N$ correlations.
This issue is also illustrated in Fig.~\ref{fig:pauli-mom-3} (c,d) by
the experimentally observed energy distributions of fragments in the $2p$ decays
of \isotope[6]{Be} and \isotope[45]{Fe}~\cite{Pfutzner:2012,Egorova:2012,
  Miernik:2007b,Grigorenko:2009}.
\emph{Qualitatively}, the signature of the Pauli focusing,
the double-hump structure characteristic for $[p^2_{3/2}]_0$ configuration,
can be easily seen in both cases.
However, it was understood (see e.g.\
Refs.~\cite{Grigorenko:2003b,Grigorenko:2009,Grigorenko:2009c,Pfutzner:2012,%
Grigorenko:2013}) that for \emph{quantitative} explanation two more important
effects require consideration:
(i) nucleon-nucleon final state interaction (FSI), and
(ii) subbarrier tunneling to configurations with lower angular momenta
(and hence easier penetration).
The $n$-$n$ FSI effect (i) leads to an enhancement of the low-energy part
of energy distribution in Fig.~\ref{fig:pauli-mom-3} (c,d).
Due to effect (ii) the following changes in the orbital populations take place
in the process of penetration through the subbarrier region:
\begin{align}
  \isotope[6]{Be}&:& [p^2_{3/2}]_0 &\to C_{23}[p^2_{3/2}]_0+C_{01}[s^2_{1/2}]_0 \, ,
                                     \label{eq:scen-6be}\\
  \isotope[45]{Fe}&:&[f^2_{7/2}]_0 &\to C_{67}[f^2_{7/2}]_0+C_{23}[p^2_{3/2}]_0 \,.
                                     \label{eq:scen-45fe}
\end{align}
The complete three-body model studies give $C_{23} \approx C_{01}$ in
\isotope[6]{Be} \cite{Grigorenko:2009c,Egorova:2012} and $C_{67}\ll C_{23}$ in
\isotope[45]{Fe} \cite{Grigorenko:2003b,Grigorenko:2009}, which provides
nice description of experimental data and explains the difference in
qualitative and quantitative descriptions of experimental spectra.

%===============================================================================

\textbf{Pauli focusing for 5-body systems.}
This effect was discussed in Ref.\ \cite{Zhukov:1994} by example of
\isotope[8]{He} nucleus described by the $\alpha$+$4n$ model.
The complicated spatial correlation patterns were predicted.
The same case was considered in more details in Ref.~\cite{Mei:2012b}.
In analogy with three-body case we may expect that Pauli focusing correlations
found in the bound five-body systems should
exhibit themselves in decays if such systems are located above the five-body
breakup threshold.
However, the questions \emph{whether} and \emph{how} such
correlations can be used to extract physical information have never been
addressed so far.

%===============================================================================

%===============================================================================
%-------------------------------------------------------------------------------
\begin{figure}[!t]
\centering
\includegraphics[width=0.224\textwidth]{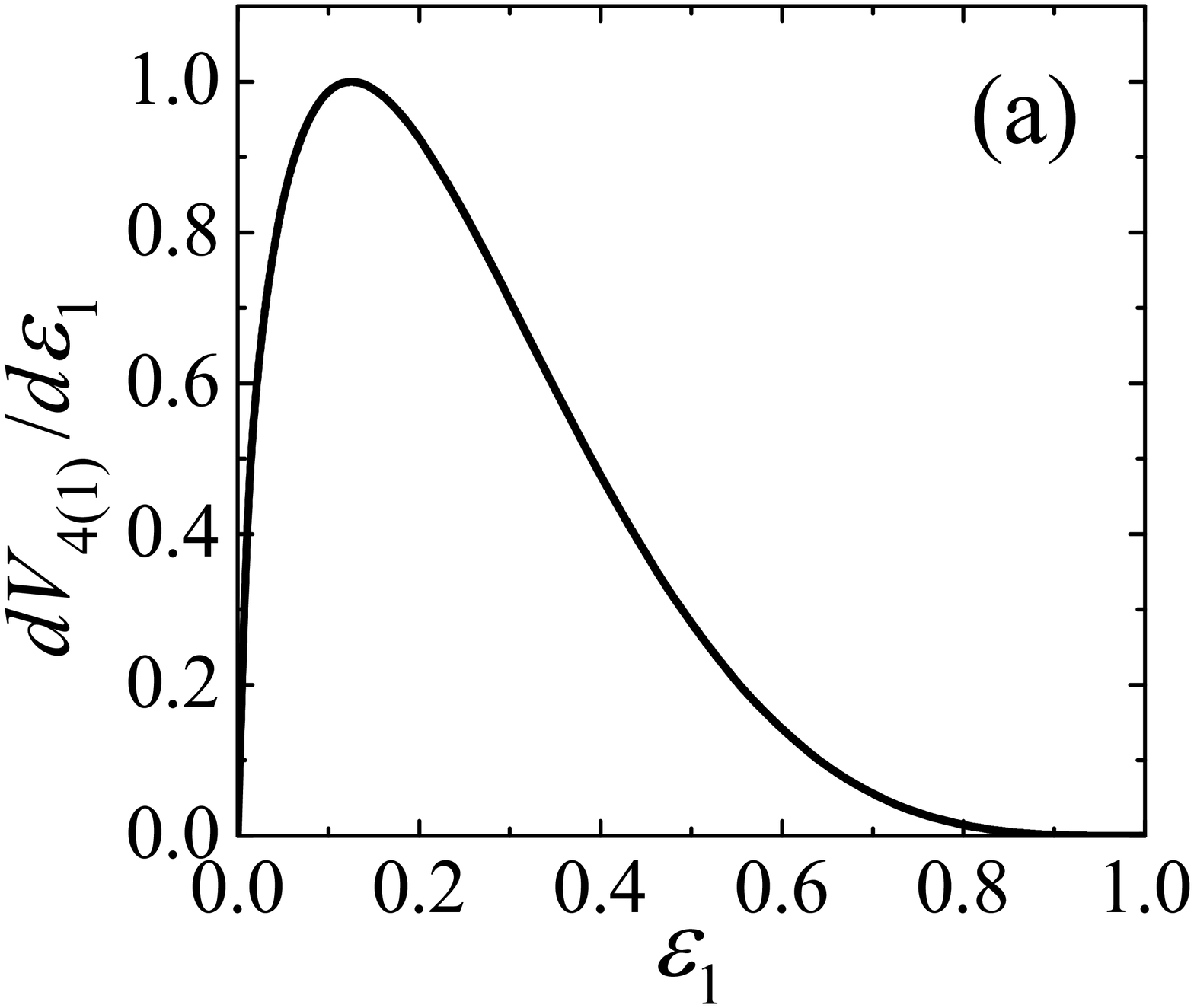}%
\includegraphics[width=0.243\textwidth]{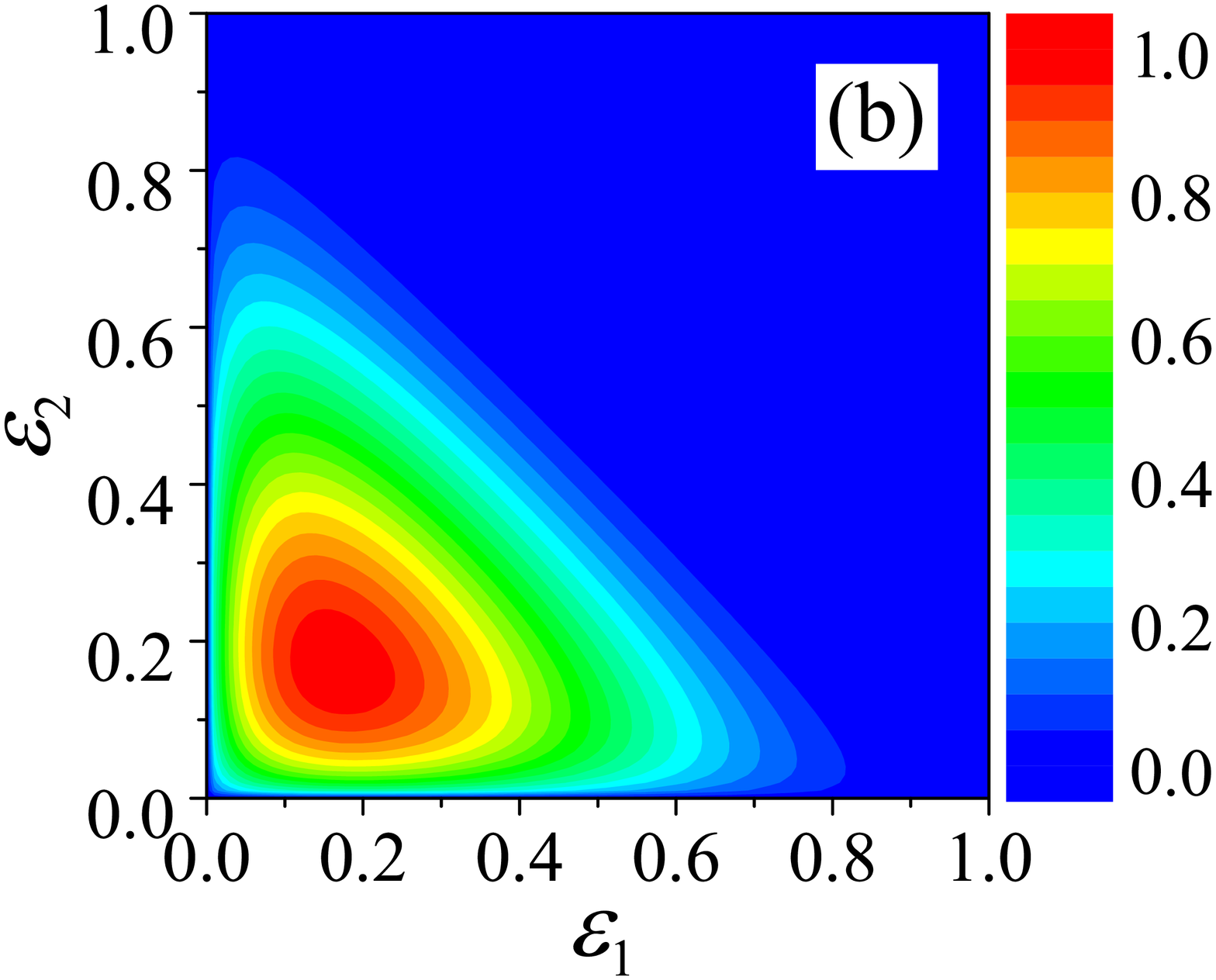}
\caption{\label{fig:dis-v1-v12} Fig.~\ref{fig:dis-v1-v12}.
  Five-body phase-space (a) one- and (b) two-dimensional energy distributions of
  Eqs.\ \eqref{eq:pw-1} and \eqref{eq:pw-12}.}
\end{figure}
%-------------------------------------------------------------------------------
\begin{figure}
\centering
\includegraphics[width=0.48\textwidth]{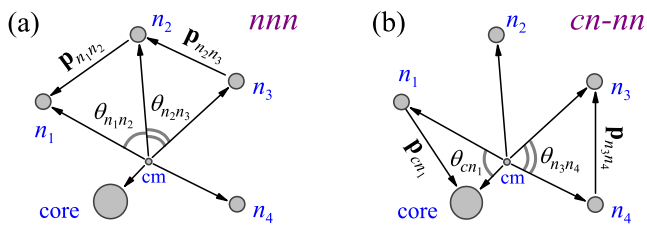}
\caption{\label{fig:scheme} Fig.~\ref{fig:scheme}.
  Schemes of kinematical variables describing 5-body decays, which are used in
  constructing correlated two-dimensional energy
  $\{\varepsilon_{ik},\varepsilon_{nm}\}$ and angular $\{\theta_{ik},\theta_{nm}\}$
  distributions of fragments.
  Examples (a) of ``connected'' $nnn$ and (b) of ``disconnected''
  $cn$-$nn$ topologies.
  The related energy distribution parameters are defined as
  $\varepsilon_{ik} = p^2_{ik}/(2\mu_{ik}E_T)$. }
\end{figure}
%-------------------------------------------------------------------------------

\textbf{Phase-space distributions.}
Before we discuss a physically meaningful signal of $4n$-decay, connected with
nuclear structure and decay dynamics, we should first understand the
``kinematically defined'' correlations, which are connected with phase-space of
several particles.
The phase-space $V_4$ for 5 particles can be defined by 4
energies $E_i=\varepsilon_i\,E_T$ corresponding to 4 Jacobi vectors in the
momentum space
\begin{equation}
  dV_4 \sim \delta(E_T-\textstyle \sum_i \varepsilon_i E_T)
  \sqrt{\varepsilon_1 \varepsilon_2  \varepsilon_3 \varepsilon_4} \,\,
  d \varepsilon_1 d \varepsilon_2  d \varepsilon_3 d \varepsilon_4 \,.
  \label{eq:pw}
\end{equation}
The dimensionless parameters $0 \leq\varepsilon_i\leq 1$ are introduced in
analogy with Eq.\ (\ref{eq:jac}).
The phase-space energy distribution between any two fragments
$dV_{4(1)}/d \varepsilon_1$ and correlated phase-space distribution of two
Jacobi energies $dV_{4(12)}/d \varepsilon_1 d \varepsilon_2$ can be obtained by
integration of Eq.~\eqref{eq:pw} over the respective variables
\begin{align}
  dV_{4(1)} &\sim \sqrt{(1-\varepsilon_1 )^7\varepsilon_1 } \,  d \varepsilon_1\,,
         \label{eq:pw-1} \\
  dV_{4(12)} &\sim (1-\varepsilon_1 - \varepsilon_2)^2
            \sqrt{\varepsilon_1 \varepsilon_2} \, d \varepsilon_1d \varepsilon_2 \,.
                   \label{eq:pw-12}
\end{align}
The distributions \eqref{eq:pw-1} and \eqref{eq:pw-12} are shown in
Fig.~\ref{fig:dis-v1-v12}(a) and (b).
The distribution \eqref{eq:pw-1} has a maximum at $\varepsilon_1 = 1/8$
and a mean energy value $\varepsilon_1 = 1/4$, the distribution \eqref{eq:pw-12}
has a maximum at $\varepsilon_1 = \varepsilon_2 =1/6$.

In addition to the two-dimensional Jacobi energy distribution of
Eq.~\eqref{eq:pw-12}, the correlated two-dimensional energy
$\{\varepsilon_{ik},\varepsilon_{nm}\}$ and angular
$\{\theta_{ik},\theta_{nm}\}$ ($i \neq k$, $n\neq m$) distributions
may be constructed, see Fig.\ \ref{fig:scheme}.
Such correlated distributions are easier to interpret in the
sense of permutation properties than the distributions for Jacobi variables.
For core+$4n$ decays in total five topologically nonequivalent
two-dimensional distributions exist.
For example, two topological types ``$nnn$'' and ``$cn$-$nn$'' are
illustrated in Fig.\ \ref{fig:scheme}.
Similarly, the types ``$ncn$'', ``$cnn$'' and ``$nn$-$nn$'' can be introduced.
The others can be reduced to them by permutations of
indistinguishable particles.

\begin{figure*}[h]
\centering
\includegraphics[width=\textwidth]{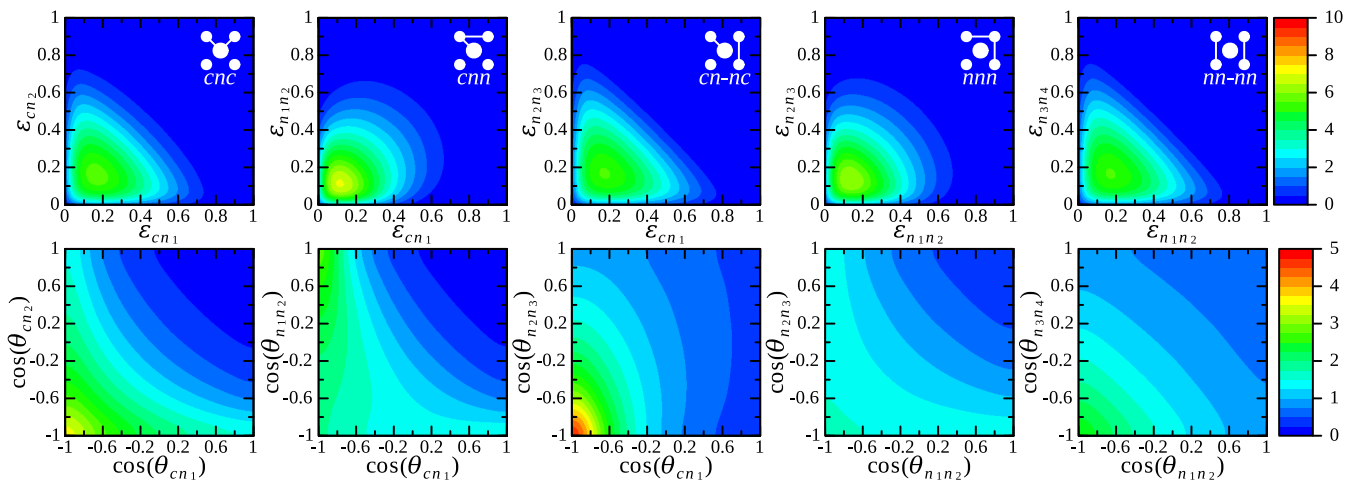}
\caption{\label{fig:dis-pw-all} Fig.~\ref{fig:dis-pw-all}.
  Nonequivalent five-body phase-space correlated energy (upper row) and angular
  (lower row) distributions for core+$4n$ decays.
  The sketches in the upper row   panels illustrate the selection of
  the correlated pairs of the fragments: $ncn$, $cnn$, $cn$-$nn$, $nnn$, $nn$-$nn$.
  For presentation purpose, the angular distributions are normalized to
  the value of 4 (the isotropic distribution then has the constant value of unity).
  The results presented in this and the
  following figures are obtained for the \isotope[7]{H} case.}
\end{figure*}

Correlated two-dimensional distributions in the phase-space case are shown in
Fig.~\ref{fig:dis-pw-all}.
The topologically nonequivalent
distributions (given in columns) are marked with sketches characterizing the
topological properties and named also by string code unambiguously related to
them.
It is evident that ``topologically disconnected'' distributions $cn$-$nn$
and $nn$-$nn$ of Fig.~\ref{fig:dis-pw-all} (cols.\ 3, 5) are equivalent to the
distributions of Eq.~\eqref{eq:pw-12} and Fig.~\ref{fig:dis-v1-v12} (b) as
Jacobi variables are ``disconnected'' in the same sense.
It can be seen that the other correlated distributions have
quite distinctive shapes even in this simple
(as there is no dynamics) phase-space case.

It should be noted that much more correlation information characterizing
core+$4n$ decay in principle exist.
However, this information can be displayed
only via correlated distributions of higher dimensions (more than two).
At the moment we find too premature to discuss such complicated things.

%-------------------------------------------------------------------------------
%-------------------------------------------------------------------------------

%===============================================================================

\textbf{Model for dynamics of 5-body decay.}
The model we develop in this work is generalization of the
\emph{improved direct $2p$-decay} model \cite{Golubkova:2016} to
the $4n$ emission case.
In direct decay models it is assumed that emitted particles are propagating to
asymptotics in fixed quantum states, while the total decay energy is shared
among single-particle configurations described by R-matrix-type amplitudes.
The differential decay probability of the 3-body decay in such a model is
\begin{align}
dW & \sim  & |T|^2 \, dV_2 \, d \Omega_1 d \Omega_2\, , \quad dV_2 = \sqrt{
\varepsilon(1-\varepsilon)} \,, \nonumber \\
T & = & \mathcal{A} \left[
A_{cn_1}(l_1,j_1,\mathbf{p}_{cn_1}) A_{cn_2}(l_2,j_2,\mathbf{p}_{cn_2})
\right]_J \,,
\label{eq:amp-anti-3}
\end{align}
where \(A_{cn_k}(l_k,j_k,\mathbf{p}_{cn_k})\) is a single particle decay amplitude
and $\mathbf{p}_{cn_k}$, $l_k$  and $j_k$ are momentum, angular momentum,
and total spin conjugated to radius-vector $\mathbf{r}_{cn_k}$ between
the core and the nucleon number $k$.
The amplitude $T$ has definite total spin \(J\) and it is antisymmetrized
(operator $\mathcal{A}$ above) for permutation of two valence nucleons.
The single-particle amplitudes are approximated by R-matrix type expression
\begin{multline}
  A_{cn_k}(l_kj_k,\mathbf{p}_{cn_k}) =
  \frac{1}{2}\frac{a_{l_kj_k}\sqrt{\Gamma_{cn_k}(E_{cn_k})}}
  {E_{r,cn_k} -E_{cn_k}-i \Gamma_{cn_k}(E_{cn_k})/2} \\
  \times  [Y_{l_k}(\hat{p}_{cn_k}) \otimes \chi_{1/2} ]_{j_k}  \,,
  \label{eq:r-matr-amp}
\end{multline}
where $E_{r,cn_k}$ is the resonance energy in $cn_k$ channel,
while $\Gamma_{cn_k}(E_{cn_k})$ is its standard  R-matrix width
as a function of the energy.
It can be found in Refs.\
\cite{Grigorenko:2007,Grigorenko:2007a} how the direct decay model stems from a
class of simplified three-body Hamiltonians and how it lead from Eqs.\
(\ref{eq:amp-anti-3}) and (\ref{eq:r-matr-amp}) to well-known R-matrix-type
expression
\begin{multline*}
\frac{dW}{d\varepsilon} = \frac{E_T(E_T-E_{r,cn_1}-E_{r,cn_2})^2}{2 \pi}\times \\
\frac{\Gamma_{cn_1}(E_{cn_1})}
{(E_{r,cn_1} -E_{cn_1})^2- \Gamma^2_{cn_1}/4} \frac{\Gamma_{cn_2}(E_{cn_2})}
{(E_{r,cn_2} -E_{cn_2})^2- \Gamma^2_{cn_2}/4} \,.
%\label{eq:prob-3}
\end{multline*}
The complete expressions of the improved direct $2p$-decay model
\cite{Golubkova:2016} taking into account effects of core recoil and angular
momentum coupling are much more complicated.

The direct decay model is powerful and reliable phenomenological tool
broadly used in the application to $2n$ and $2p$ decays for lifetime estimates
\cite{Azhari:1998,Brown:2003,Barker:2003,Olsen:2013,Brown:2015},
studies of two-nucleon correlations \cite{Pfutzner:2012,Golubkova:2016} and
transitional dynamics \cite{Mukha:2015,Golubkova:2016,Xu:2018}.
Special efforts were made to validate
\cite{Grigorenko:2007,Grigorenko:2007a,Pfutzner:2012}
and to improve the model \cite{Golubkova:2016}.

Following the approach of Eq.\ (\ref{eq:amp-anti-3}) the differential decay
probability of the 5-body decay for $\mathrm{core}+n_1+n_2+n_3+n_4$ system is
\begin{align}
dW & \sim  |T|^2 \, dV_4 \, \textstyle \prod_{k=1..4} d \Omega_k \, , \nonumber
\\
T & =  \mathcal{A} \left[ \textstyle \prod_{k=1..4}
A_{cn_k}(l_k,j_k,\mathbf{p}_{cn_k})\right]_J.
\label{eq:amp-anti}
\end{align}
The amplitude $T$ has defined total spin \(J\) and is antisymmetrized for
permutation of four valence nucleons.
Simplified version of the formalism
allowing the compact analytical expressions was applied in Ref.\
\cite{Grigorenko:2011} to estimates of $4n$ decay widths.
In this work we use no
simplifications and therefore employ the Monte-Carlo procedure to compute
various momentum distributions for full version of the formalism.

%-------------------------------------------------------------------------------
\begin{figure}[!h]
\centering
\includegraphics[width=0.24\textwidth]{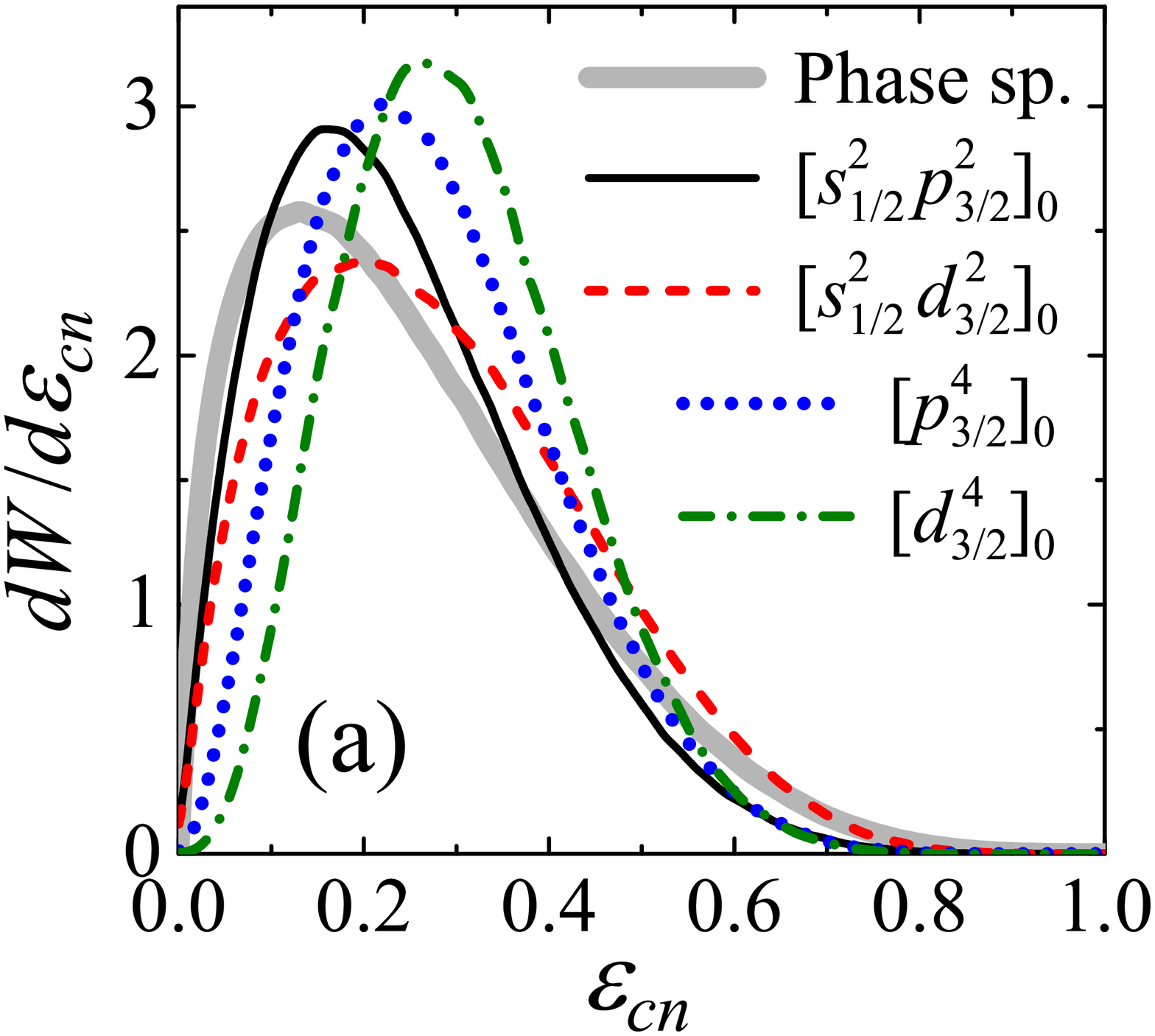}%
\includegraphics[width=0.24\textwidth]{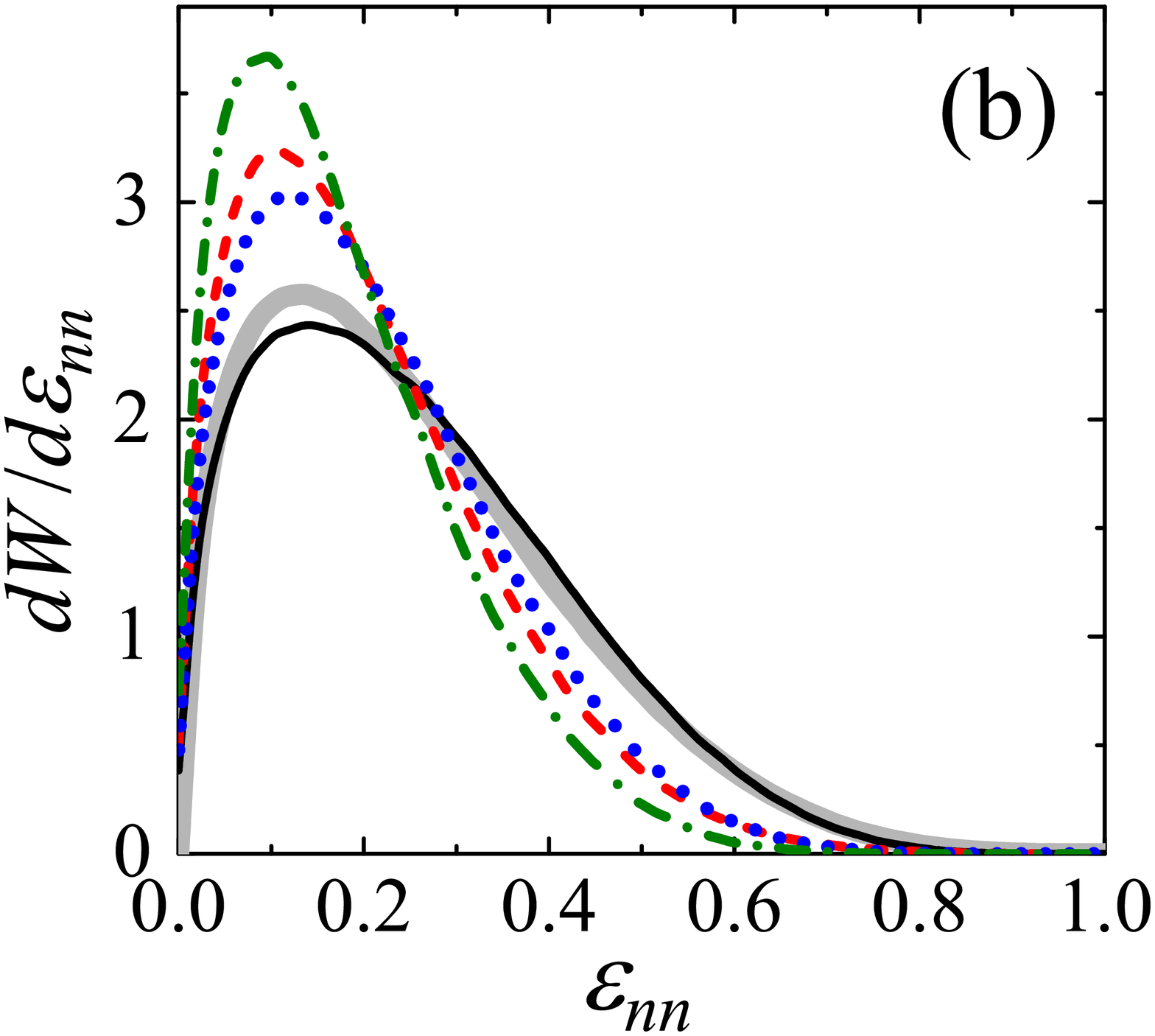}
\includegraphics[width=0.24\textwidth]{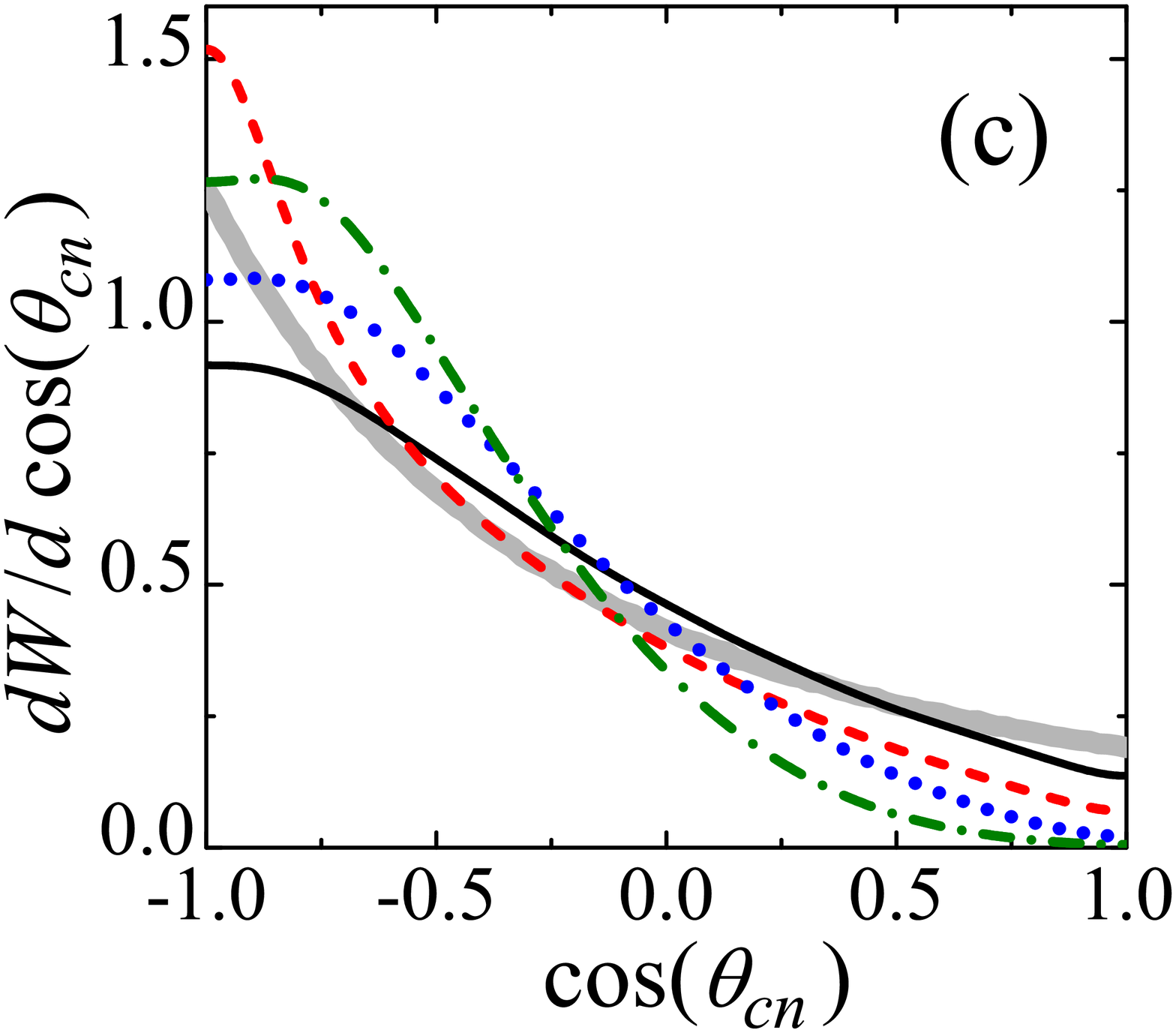}%
\includegraphics[width=0.24\textwidth]{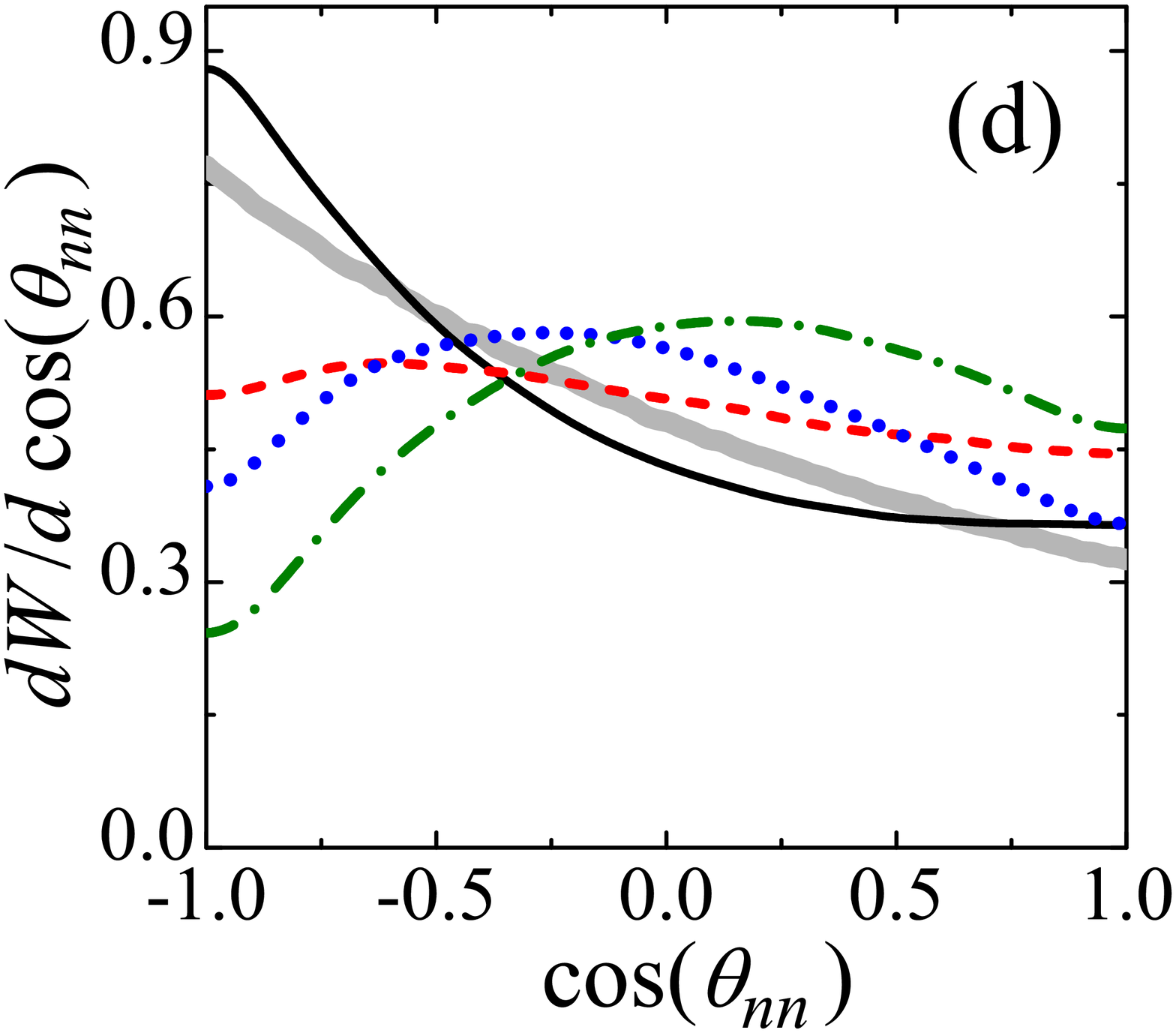}
\caption{\label{fig:en-ang-all} Fig.~\ref{fig:en-ang-all}.
  The normalized energy (a,b) and angular (c,d) distributions in the core-$n$
  (a,c) and $n$-$n$ (b,d) channels of $4n$-decay for all different pure $4n$
  configurations which can be found in Eqs.\ \eqref{eq:scen-7h} and
  \eqref{eq:scen-28o}.
  The decay energy is \(E_T=500\) keV.
  The reference phase-space distributions
  are shown by the thick gray lines.}
\end{figure}
%-------------------------------------------------------------------------------

%-------------------------------------------------------------------------------
\begin{figure*}[!h]
\centering
\includegraphics[width=0.75\textwidth]{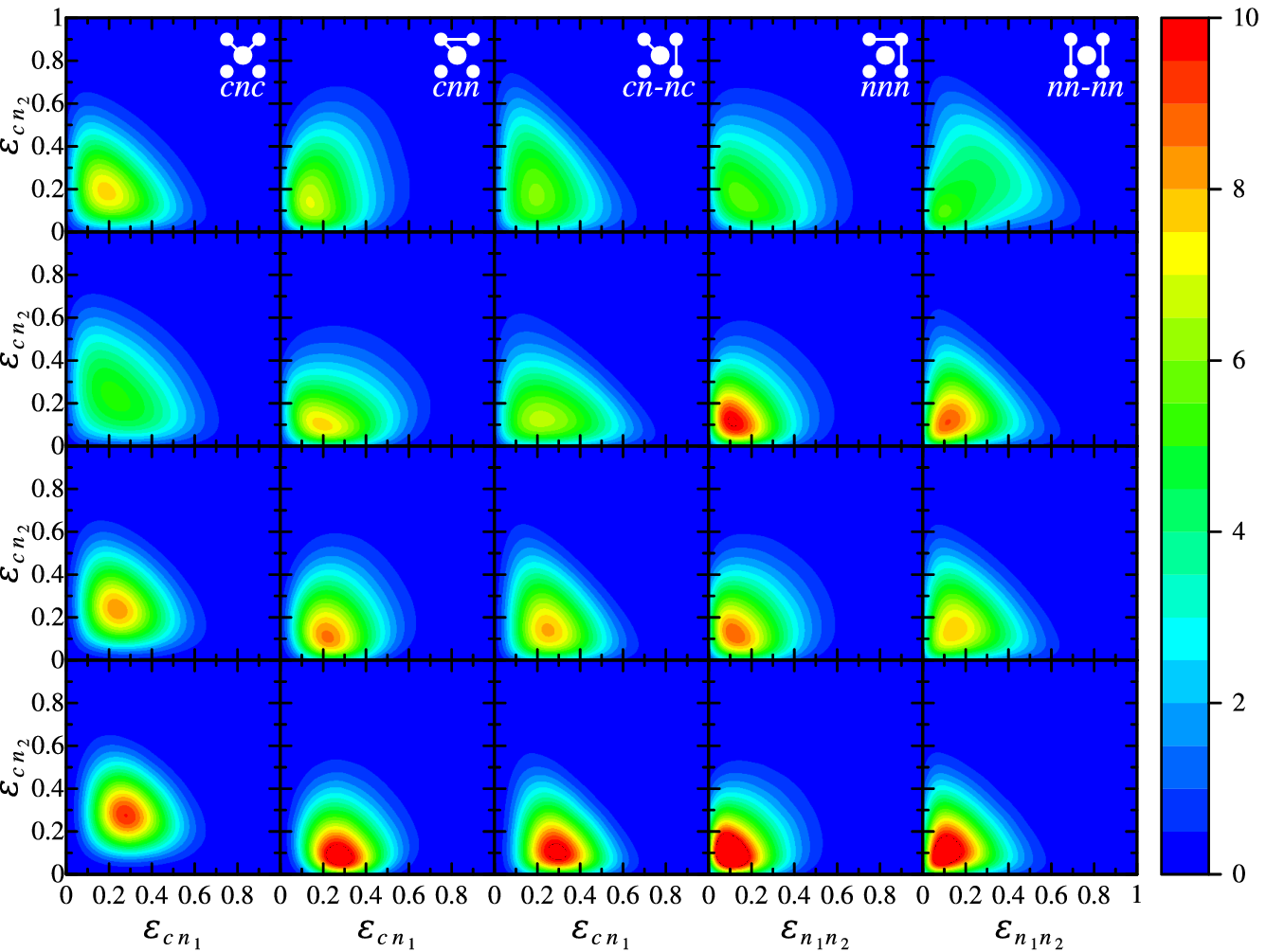}%
\caption{\label{fig:corel-eps-all} Fig.~\ref{fig:corel-eps-all}.
  The correlated energy distributions for $ncn$ (col.\ 1), $cnn$ (col.\ 2),
  $cn$-$nn$ (col.\ 3), $nnn$ (col.\ 4), and $nn$-$nn$ (col.\ 5) topologies.
  Configurations from Eqs.\ \eqref{eq:scen-7h} and \eqref{eq:scen-28o}
  correspond to the pure components $[s^2_{1/2} p^2_{3/2}]_0$ (row 1),
  $[s^2_{1/2} d^2_{3/2}]_0$ (row 2), $[p^4_{3/2}]_0$ (row 3),
  and $[d^4_{3/2}]_0$ (row 4).}
\end{figure*}
%-------------------------------------------------------------------------------

\begin{figure}[b]
\centering
\includegraphics[width=0.24\textwidth]{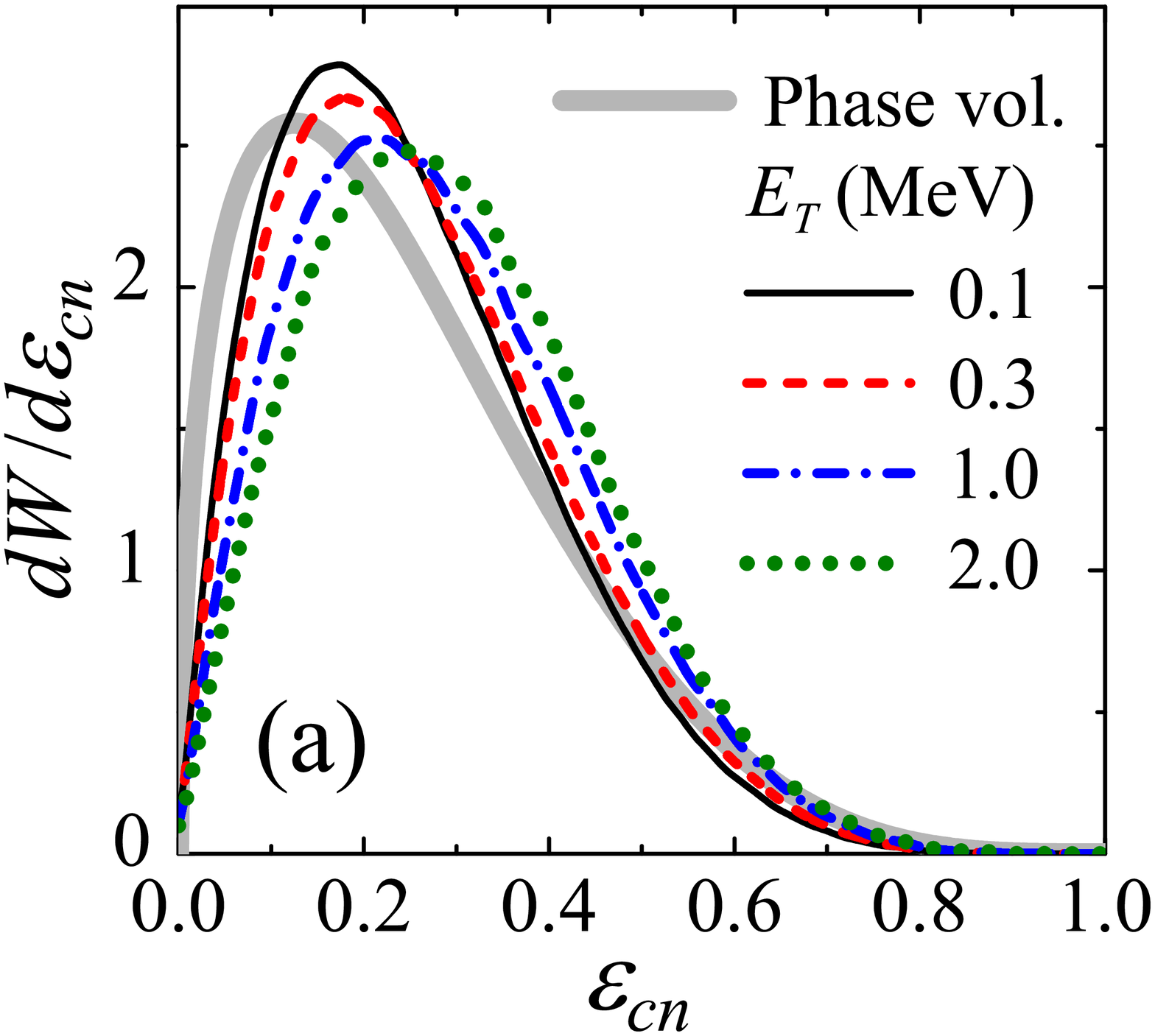}%
\includegraphics[width=0.24\textwidth]{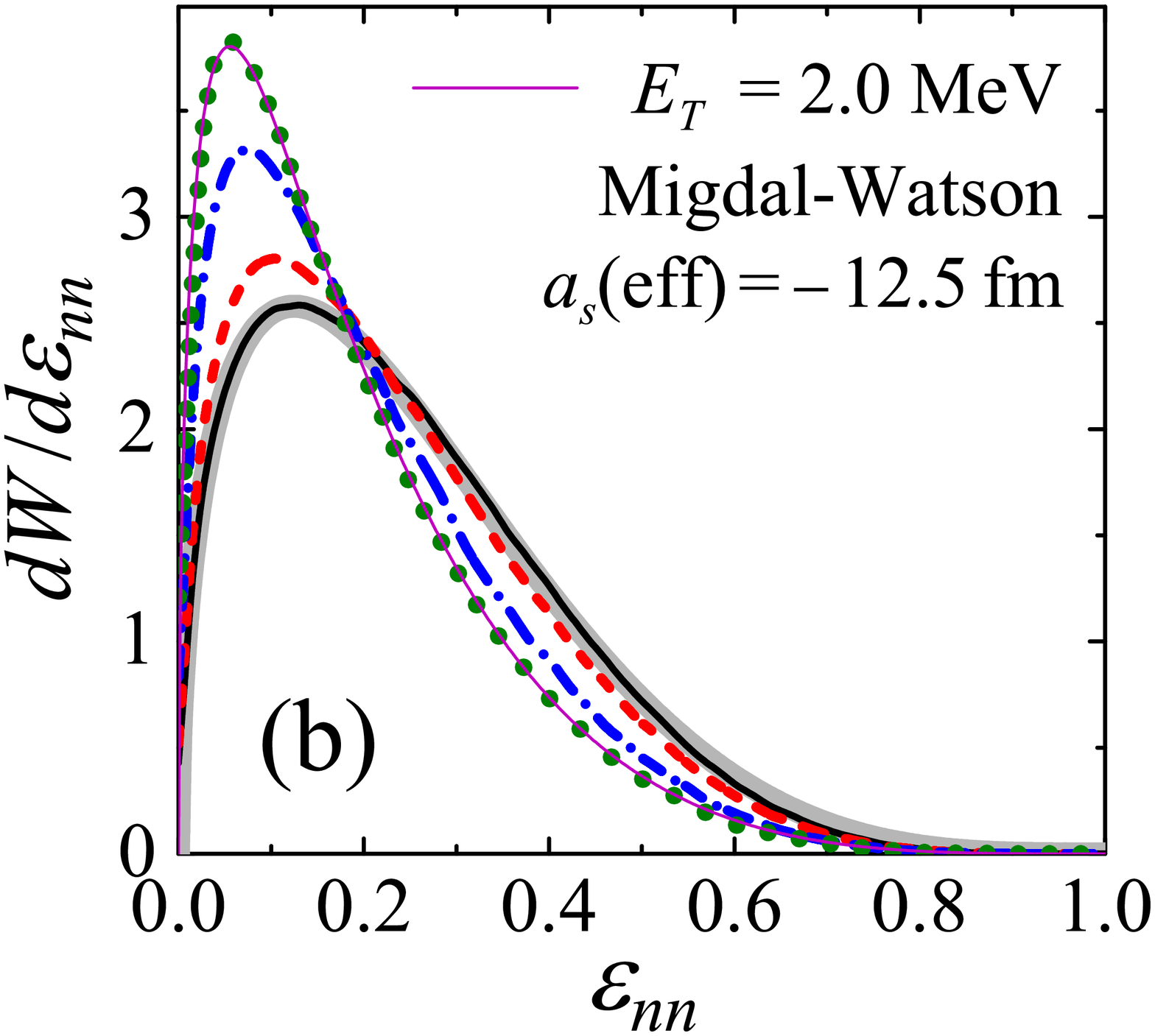}
\caption{\label{fig:nn-fsi-all} Fig.~\ref{fig:nn-fsi-all}.
  Effect of $n$-$n$ FSI on energy distributions in core-$n$ (a) and $n$-$n$ (b)
  channels at different decay energies $E_T$ for $[s^2_{1/2}p^2_{3/2}]$
  configuration.
  Solid, dashed, dash-dotted, and dotted lines correspond to
  \(E_T\) of 0.1, 0.3, 1.0, and 2.0 MeV, respectively.
  Phase-space is shown by thick grey line.
  Thin solid line in (b) shows the results of Eq.\ (\ref{eq:mw})
  for \(E_T=2\) MeV.}
\end{figure}

In order to estimate the possible effect of $n$-$n$ FSI, the
Eq.~\eqref{eq:amp-anti} can be modified as
\begin{equation}
  T \!=\!\mathcal{A}\!\!\left[ \textstyle\prod_{k=1..4}A_{cn_k}(\mathbf{p}_{cn_k})
    \!\textstyle \prod_{m > k} A_{n_kn_m}(\mathbf{p}_{n_kn_m})\right]\!.
  \label{eq:amp-anti-nn}
\end{equation}
The amplitude $A_{n_kn_m}$ is the
(typical for the MW-type approximations) zero-range expression
for an $s$-wave $n$-$n$ scattering
\begin{equation}
  A_{n_kn_m} = \frac{a_s}{1-i\, p_{n_k n_m}\, a_s} \, ,
  \label{eq:amp-mw}
\end{equation}
where $a_s=-18.9$ fm is the singlet scattering length in the $n$-$n$ channel.
This approximation provides no control of the actual total spin in the
particular $n_k$-$n_m$ channel.
Thus the effect of $n$-$n$ FSI should be much overestimated here as,
\emph{on average}, only two pairs of neutrons (out of six possible selections)
can be both in relative spin \(S=0\) state and in relative $s$-wave simultaneously.
So, Eq.~\eqref{eq:amp-anti-nn} cannot be consistently used for evaluation of
the $n$-$n$ FSI effect on momentum distributions.
However, it can be used to provide a reliable upper-limit estimate of the effect,
see Fig.\ \ref{fig:nn-fsi-all} and the related text.

%===============================================================================

\textbf{Scenarios of 5-body decay.}
The five-body decays are so far not studied at all.
Thus, in our discussion we follow the assumed qualitative analogy of 3-body
and 5-body decay dynamics.
Let us consider the examples of $^{7}$H and $^{28}$O whose internal structure
is expected to be dominated by $[p^4_{3/2}]_0$ and
$[d^4_{3/2}]_0$ configurations, respectively.
In the decay process these configurations may drift to the configurations
with lower angular momenta in analogy with
Eqs.\ \eqref{eq:scen-6be} and \eqref{eq:scen-45fe}
\begin{align}
  \isotope[7]{H} : [p^4_{3/2}]_0 \to C_{2323} [p^4_{3/2}]_0
  &+  C_{0123} [s^2_{1/2} p^2_{3/2}]_0, \label{eq:scen-7h} \\
  \isotope[28]{O}: [d^4_{3/2}]_0 \to C_{4343} [d^4_{3/2}]_0
  &+  C_{0143} [s^2_{1/2} d^2_{3/2}]_0 \phantom{.} \nonumber  \\
  &+  C_{0123} [s^2_{1/2} p^2_{3/2}]_0. \label{eq:scen-28o}
\end{align}
The one-dimensional energy and angular distributions for all the pure
above-mentioned configurations in the core-$n$ and $n$-$n$ channels
are shown in Fig.~\ref{fig:en-ang-all}.
On average, the increase of angular momentum ``content'' of the component leads
to the increase of the mean energy in the core-$n$ and
the corresponding decrease in the $n$-$n$ channels.
The respective angular distributions have quite distinctive patterns.
So, one may see a strong effect of the definite orbital configuration already
in these observables.

\begin{figure*}[t]
\centering
\includegraphics[width=0.75\textwidth]{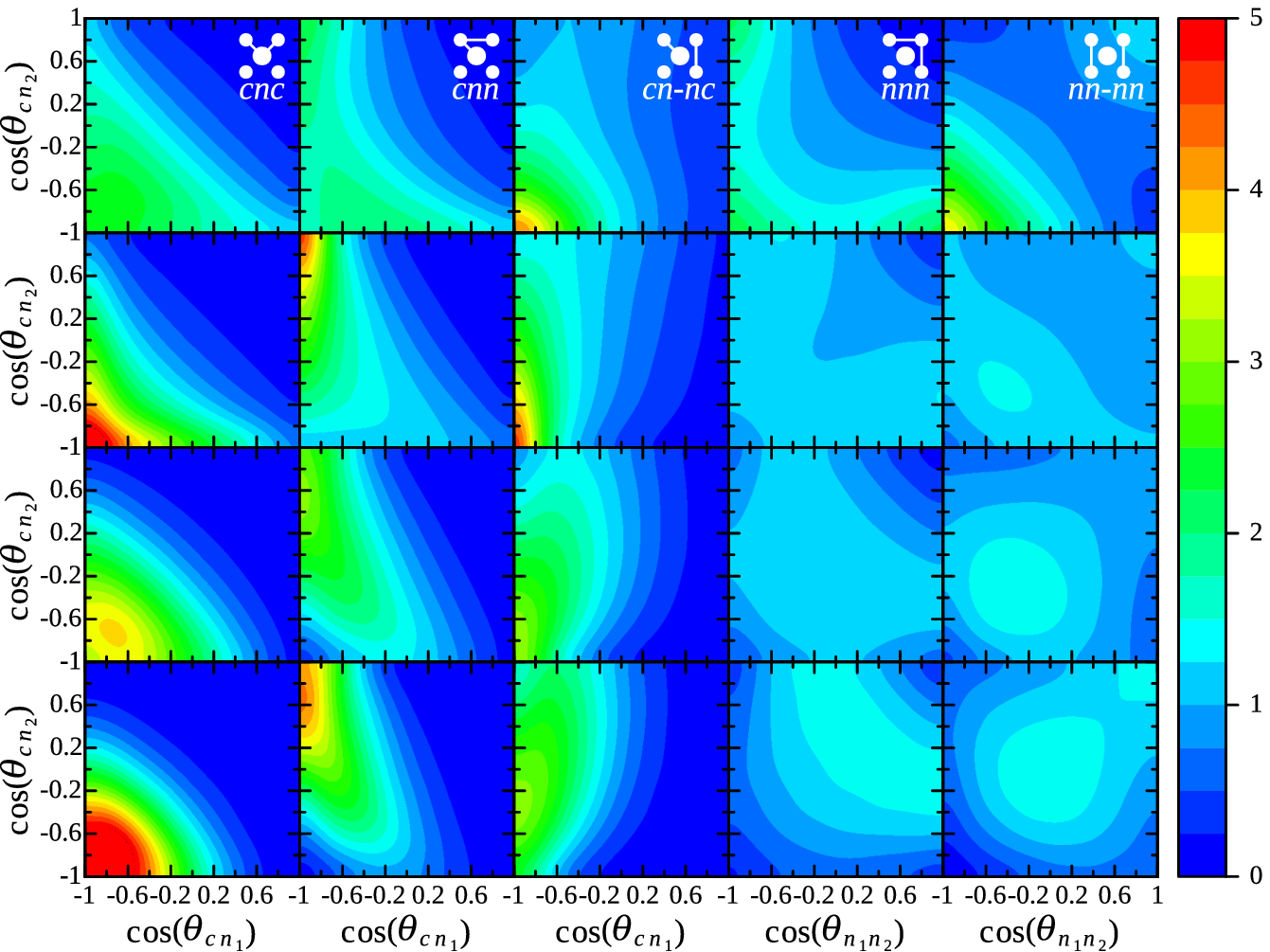}
\caption{\label{fig:corel-ang-all} Fig.~\ref{fig:corel-ang-all}.
  The correlated angular distributions.
  Panels layout is the same as in the Fig.\ \ref{fig:corel-eps-all}.}
\end{figure*}

The Pauli focusing effect on two-dimensional correlated energy distributions
for pure configurations from Eqs.\ (\ref{eq:scen-7h}) and (\ref{eq:scen-28o})
is illustrated in Fig.~\ref{fig:corel-eps-all}.
All four considered cases are quite distinguishable from phase-space
distributions of Fig.\ \ref{fig:dis-pw-all}, as well as from each other.
Among them the case of $[s^2_{1/2} p^2_{3/2}]_0$ configuration has
especially strong qualitative distinction.
Even stronger variability can be seen in the correlated angular distributions,
which are systematically surveyed in Fig.~\ref{fig:corel-ang-all}.
%-------------------------------------------------------------------------------
%-------------------------------------------------------------------------------
The correlation patterns show great variability: all of them are
strongly different from phase-space distributions and from each other.
Therefore, the full set of predicted correlation patterns provides
clear signatures on dominating $4n$-decay configurations:
some pairs of plots for two different configurations may look similar, however,
a selection of other pair should provide a distinctive signal.
This feature may be used in planning of future experiments and the data
analysis.

%===============================================================================

\textbf{On applicability of the model.}
The experience of three-body decay studies tells us that the effect of
configuration mixing in the final state may have strong effect on the observed correlations.
However, the simplified predicted correlations may be relevant in two limiting cases.
(i) For decay energies high enough above the typical barrier energies
(e.g.\ $E_T>2-4$ MeV)
the effect of configuration mixing may be small,
and then one-configuration approximation may be precise enough.
However, $n$-$n$ FSI should have growing effect on the distributions in such a limiting case.
(ii) For deep subbarrier energies $E_T<100-300$ keV, the decay should proceed
via the $[s^2p^2]_0$ configurations which have the lowest possible barriers.
For the case (ii) our predictions should be very precise with improving
precision in the limit $E_T \rightarrow 0$.
Below we elaborate more on the
latter ``reliable mode'' of the model application.

%-------------------------------------------------------------------------------

%-------------------------------------------------------------------------------
\begin{figure*}[!ht]
\centering
\includegraphics[width=0.75\textwidth]{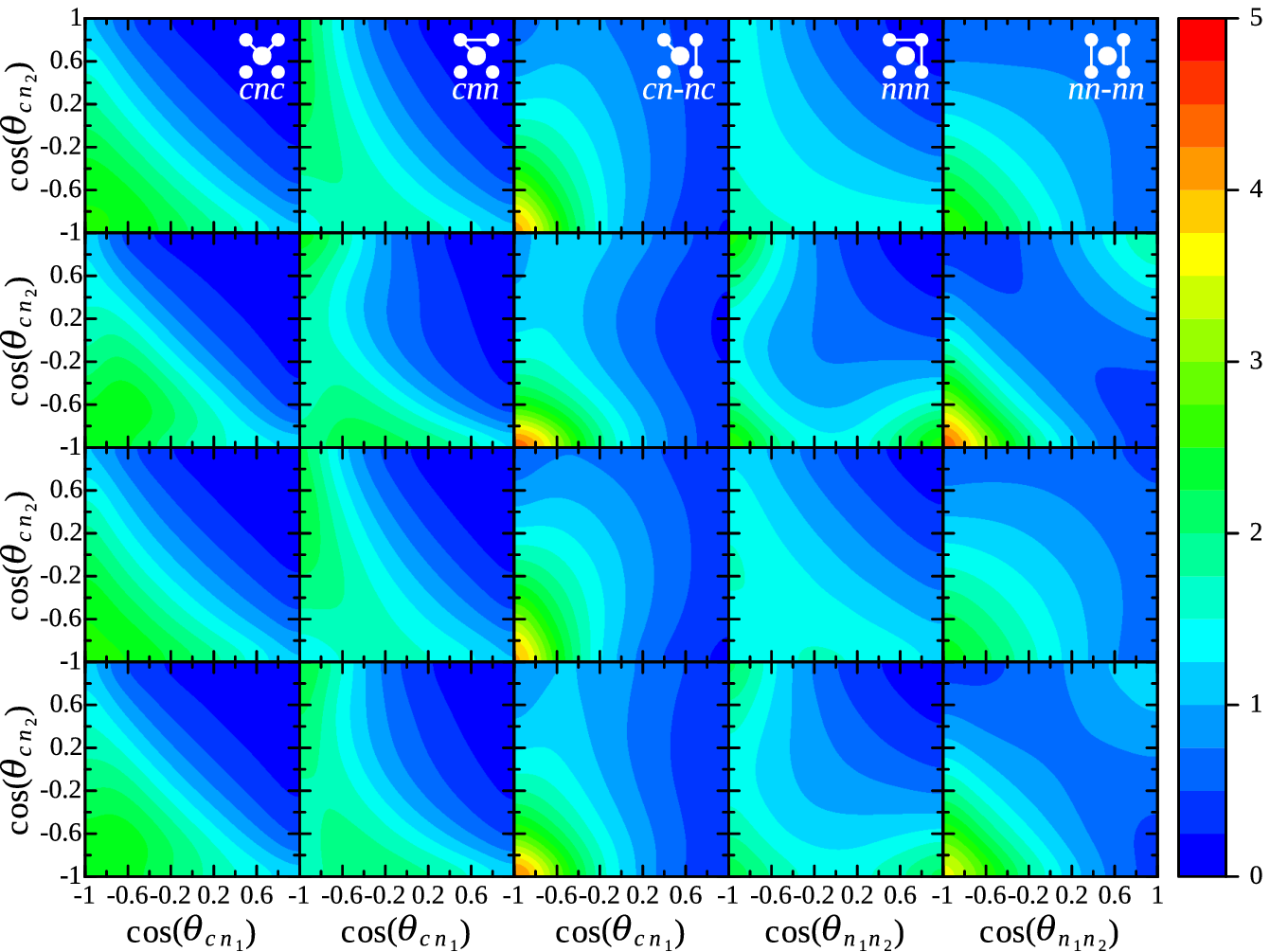}
\caption{\label{fig:p12-p32-mix}  Fig.~\ref{fig:p12-p32-mix}.
Correlated angular distributions for core+$4n$ decays in the case of
$C_{0123}[s^2_{1/2}p^2_{3/2}]_0+C_{0121}[s^2_{1/2}p^2_{1/2}]_0$ configuration
mixing for different topologies (columns). The following mixing cases are
illustrated in the rows: (row 1) $C_{0121}=1$ (pure $[s^2_{1/2}p^2_{1/2}]_0$),
(row 2) $C_{0121}=0.48$, (row 3) $C_{0121}=-0.48$ (both have $24\%$ of
$[s^2_{1/2}p^2_{1/2}]_0$), and (row 4) $C_{0121}=0$ (pure
$[s^2_{1/2}p^2_{3/2}]_0$).}
\end{figure*}
%-------------------------------------------------------------------------------

%===============================================================================

\textbf{Low-energy limit: Effect of $n$-$n$ FSI.}
This effect is known to influence strongly the three-body energy and angular
correlations.
In three-body case this effect is expected to vanish as the decay
energy goes below $E_T \sim 100$ keV, see, e.g.~\cite{Grigorenko:2018}.
It could be important to find out when this effect becomes negligible for
$4n$ emission.

Distributions of Fig.~\ref{fig:nn-fsi-all} obtained by
Eq.~\eqref{eq:amp-anti-nn} for $[s^2_{1/2}p^2_{3/2}]_0$ configuration illustrate
how important can be the effect of $n$-$n$ FSI in five-body decay.
%-------------------------------------------------------------------------------
We should emphasize here that this is the upper-limit estimate of the effect,
see discussion around Eq.\ (\ref{eq:amp-anti-nn}), and actual scale of the effect
can be only smaller.
We can see in Fig.~\ref{fig:nn-fsi-all}~(b) that the effect is reasonably small
under $E_T<300$ keV and can be neglected under $E_T<100$ keV.
It is curious to note that the modifications of the $n$-$n$ energy distribution
by the $n$-$n$ FSI for $E_T \lesssim 2$ MeV is near perfectly described by
the MW-type expression
\begin{equation}
  \dfrac{d W}{d \varepsilon_{nn}} = \dfrac{d W_{ps}}{d \varepsilon_{nn}} \,
  \dfrac{a_s^2(\mathrm{eff})}{1+2M_{nn} \, E_T \, \varepsilon_{nn} \,
  a_s^2(\mathrm{eff})} \,,
  \label{eq:mw}
\end{equation}
where $d W_{ps}/d \varepsilon_{nn}$ is the phase-space distribution and
$a_s(\mathrm{eff})=-12.5$ fm is a kind of the effective scattering length,
see the thin solid curve in Fig.~\ref{fig:nn-fsi-all} (b), which practically
coincides with $E_T=2$ MeV results.
Thus we can qualitatively state that the ``effective intensity'' of the $n$-$n$
FSI within such a ``tetraneutron'' (namely, $[s^2_{1/2}p^2_{3/2}]_0$) is around
$44 \%$ of the ``vacuum intensity'' evaluated as \(a_s^2\) (with $a_s=-18.9$
fm).

%===============================================================================

\textbf{Low-energy limit: $p_{1/2}$ and $p_{3/2}$ mixing.}
As we have shown above for the low-energy $4n$ decays effects of the $n$-$n$ FSI
vanishes.
Only $[s^2p^2]_0$ with the smallest centrifugal barriers can be ``active'' here.
For the decay, e.g.\ of the $^{7}$H system with closed $p_{3/2}$
subshell, it is very reasonable to expect strong domination of
$[s^2_{1/2}p^2_{3/2}]_0$ configuration.
However, we cannot exclude some mixing with $[s^2_{1/2}p^2_{1/2}]_0$
configuration.
This mixing remains the only potentially uncertain aspect of the model in the
low-energy limit.

Figure \ref{fig:p12-p32-mix} shows that the sets of the predicted correlation
patterns are not only sufficiently different for the pure
$[s^2_{1/2}p^2_{3/2}]_0$ and $[s^2_{1/2}p^2_{1/2}]_0$ configurations.
Even for the moderate admixture of $p_{1/2}$ configuration on the level
$10-25 \%$ the correlation patterns are sensitive enough to clearly distinguish
the mixing ratio and the interference sign.
So, in the specific situation of the low-energy decay, studies of
the Pauli focusing correlations should be able provide basis for extraction of
quite detailed and precise information about decay dynamics.
To mention again that in this situation the predictions of Fig.\
\ref{fig:p12-p32-mix} are reliable in the sense that they are actually free from
theoretical model assumptions.

%===============================================================================

\textbf{Conclusion.}
In this work we have for the first time theoretically studied the correlations
in emission of four nucleons in the nuclear 5-body decay.
We have demonstrated that for true five-body decays of core+$4n$ systems
the \emph{Pauli focusing} --- the cumulative effects of antisymmetrization and
population of definite orbital configurations ---
may lead to distinctive correlation patterns.
These patterns are not very expressed in the one-dimensional distributions.
Here they can also be masked by other dynamical effects.
For that reason we propose to study the
\emph{full set of the two-dimensional correlated energy or/and angular distributions}
for derivation of the information concerning the quantum-mechanical $4n$-decay configuration.
In total five non-equivalent correlated distributions are available for analysis
and taken together they are predicted to form a unique ``fingerprint'' of the
decaying quantum state.
The reconstruction of all these distributions requires a complete kinematical
characterization of the core+$4n$ decay, which is presumably within
the reach of the modern experiment.

We emphasize that in general case the results of the presented model can be seen
only as starting point for qualitative discussion.
However, for the limiting case of low-energy decays (e.g.\ $E_T<300$ keV)
the model provides absolutely reliable predictions (e.g.\ concerning the mixing
of $[s^2_{1/2}p^2_{3/2}]_0$ and $[s^2_{1/2}p^2_{1/2}]_0$ configurations).

%===============================================================================
%
\textbf{Acknowledgments.}
This work for P. G. Sharov and L. V. Grigorenko was supported in part by
the Russian Science Foundation grant No.\ 17-12-01367.
The authors are grateful to A.\ Fomichev, I.\ Mukha, E.Yu.\ Nikolskii,
and G.M.\ Ter-Akopian for helpful discussions.

\bibliographystyle{ieeetr}
\bibliography{all}

\begin{thebibliography}{10}

\bibitem{Pfutzner:2012}
M.~Pf\"utzner, M.~Karny, L.~V. Grigorenko, and K.~Riisager, ``Radioactive
  decays at limits of nuclear stability,'' {\em Rev. Mod. Phys.}, vol.~84,
  pp.~567--619, Apr 2012.

\bibitem{Grigorenko:2003b}
L.~V. Grigorenko and M.~V. Zhukov, ``Two-proton radioactivity and three-body
  decay. ii. exploratory studies of lifetimes and correlations,'' {\em Phys.
  Rev. C}, vol.~68, p.~054005, 2003.

\bibitem{Mukha:2006}
I.~Mukha, E.~Roeckl, L.~Batist, A.~Blazhev, J.~D\"{o}ring, H.~Grawe,
  L.~Grigorenko, M.~Huyse, Z.~Janas, R.~Kirchner, M.~L. Commara, C.~Mazzocchi,
  S.~L. Tabor, and P.~V. Duppen, ``Proton-proton correlations observed in
  two-proton radioactivity of $^{94}$ag,'' {\em Nature}, vol.~439,
  pp.~298--302, 2006.

\bibitem{Grigorenko:2007}
L.~V. Grigorenko and M.~V. Zhukov, ``Two-proton radioactivity and three-body
  decay. iii. integral formulas for decay widths in a simplified semianalytical
  approach,'' {\em Phys. Rev. C}, vol.~76, p.~014008, Jul 2007.

\bibitem{Grigorenko:2007a}
L.~V. Grigorenko and M.~V. Zhukov, ``Two-proton radioactivity and three-body
  decay. iv. connection to quasiclassical formulation,'' {\em Phys. Rev. C},
  vol.~76, p.~014009, 2007.

\bibitem{Mukha:2008}
I.~Mukha, L.~Grigorenko, K.~S\"ummerer, L.~Acosta, M.~A.~G. Alvarez,
  E.~Casarejos, A.~Chatillon, D.~Cortina-Gil, J.~M. Espino, A.~Fomichev, J.~E.
  Garc\'ia-Ramos, H.~Geissel, J.~G\'omez-Camacho, J.~Hofmann, O.~Kiselev,
  A.~Korsheninnikov, N.~Kurz, Y.~Litvinov, I.~Martel, C.~Nociforo, W.~Ott,
  M.~Pf\"utzner, C.~Rodr\'iguez-Tajes, E.~Roeckl, M.~Stanoiu, H.~Weick, and
  P.~J. Woods, ``Proton-proton correlations observed in two-proton decay of
  $^{19}\mathrm{Mg}$ and $^{16}\mathrm{Ne}$,'' {\em Phys. Rev. C}, vol.~77,
  p.~061303, Jun 2008.

\bibitem{Grigorenko:2009}
L.~V. Grigorenko, T.~D. Wiser, K.~Miernik, R.~J. Charity, M.~Pf\"utzner,
  A.~Banu, C.~R. Bingham, M.~Cwiok, I.~G. Darby, W.~Dominik, J.~M. Elson,
  T.~Ginter, R.~Grzywacz, Z.~Janas, M.~Karny, A.~Korgul, S.~N. Liddick,
  K.~Mercurio, M.~Rajabali, K.~Rykaczewski, R.~Shane, L.~G. Sobotka, A.~Stolz,
  L.~Trache, R.~E. Tribble, A.~H. Wuosmaa, and M.~V. Zhukov, ``Complete
  correlation studies of two-proton decays: $^{6}$be and $^{45}$fe,'' {\em
  Phys. Lett. B}, vol.~677, pp.~30--35, 2009.

\bibitem{Grigorenko:2009c}
L.~V. Grigorenko, T.~D. Wiser, K.~Mercurio, R.~J. Charity, R.~Shane, L.~G.
  Sobotka, J.~M. Elson, A.~H. Wuosmaa, A.~Banu, M.~McCleskey, L.~Trache, R.~E.
  Tribble, and M.~V. Zhukov, ``Three-body decay of $^6$be,'' {\em Phys. Rev.
  C}, vol.~80, p.~034602, Sep 2009.

\bibitem{Egorova:2012}
I.~A. Egorova, R.~J. Charity, L.~V. Grigorenko, Z.~Chajecki, D.~Coupland, J.~M.
  Elson, T.~K. Ghosh, M.~E. Howard, H.~Iwasaki, M.~Kilburn, J.~Lee, W.~G.
  Lynch, J.~Manfredi, S.~T. Marley, A.~Sanetullaev, R.~Shane, D.~V. Shetty,
  L.~G. Sobotka, M.~B. Tsang, J.~Winkelbauer, A.~H. Wuosmaa, M.~Youngs, and
  M.~V. Zhukov, ``Democratic decay of {Be}6 exposed by correlations,'' {\em
  Phys. Rev. Lett.}, vol.~109, p.~202502, Nov 2012.

\bibitem{Grigorenko:2013}
L.~V. Grigorenko, I.~G. Mukha, and M.~V. Zhukov, ``Lifetime and fragment
  correlations for the two-neutron decay of $^{26}\mathbf{O}$ ground state,''
  {\em Phys. Rev. Lett.}, vol.~111, p.~042501, Jul 2013.

\bibitem{Brown:2014}
K.~W. Brown, R.~J. Charity, L.~G. Sobotka, Z.~Chajecki, L.~V. Grigorenko, I.~A.
  Egorova, Y.~L. Parfenova, M.~V. Zhukov, S.~Bedoor, W.~W. Buhro, J.~M. Elson,
  W.~G. Lynch, J.~Manfredi, D.~G. McNeel, W.~Reviol, R.~Shane, R.~H. Showalter,
  M.~B. Tsang, J.~R. Winkelbauer, and A.~H. Wuosmaa, ``Observation of
  long-range three-body coulomb effects in the decay of $^{16}\mathrm{Ne}$,''
  {\em Phys. Rev. Lett.}, vol.~113, p.~232501, Dec 2014.

\bibitem{Brown:2015}
K.~W. Brown, R.~J. Charity, L.~G. Sobotka, L.~V. Grigorenko, T.~A. Golubkova,
  S.~Bedoor, W.~W. Buhro, Z.~Chajecki, J.~M. Elson, W.~G. Lynch, J.~Manfredi,
  D.~G. McNeel, W.~Reviol, R.~Shane, R.~H. Showalter, M.~B. Tsang, J.~R.
  Winkelbauer, and A.~H. Wuosmaa, ``Interplay between sequential and prompt
  two-proton decay from the first excited state of $^{16}\mathrm{Ne}$,'' {\em
  Phys. Rev. C}, vol.~92, p.~034329, Sep 2015.

\bibitem{Grigorenko:2018}
L.~V. Grigorenko, J.~S. Vaagen, and M.~V. Zhukov, ``Exploring the manifestation
  and nature of a dineutron in two-neutron emission using a dynamical dineutron
  model,'' {\em Phys. Rev. C}, vol.~97, p.~034605, Mar 2018.

\bibitem{Grigorenko:2011}
L.~V. Grigorenko, I.~G. Mukha, C.~Scheidenberger, and M.~V. Zhukov, ``On
  two-neutron radioactivity and four-nucleon emission of exotic nuclei,'' {\em
  Phys. Rev. C}, vol.~84, p.~021303(R), 2011.

\bibitem{Charity:2011}
R.~J. Charity, J.~M. Elson, J.~Manfredi, R.~Shane, L.~G. Sobotka, B.~A. Brown,
  Z.~Chajecki, D.~Coupland, H.~Iwasaki, M.~Kilburn, J.~Lee, W.~G. Lynch,
  A.~Sanetullaev, M.~B. Tsang, J.~Winkelbauer, M.~Youngs, S.~T. Marley, D.~V.
  Shetty, A.~H. Wuosmaa, T.~K. Ghosh, and M.~E. Howard, ``Investigations of
  three-, four-, and five-particle decay channels of levels in light nuclei
  created using a c9 beam,'' {\em Phys. Rev. C}, vol.~84, p.~014320, Jul 2011.

\bibitem{Migdal:1955}
A.~Migdal, ``The theory of nuclear reactions with production of slow
  particles,'' {\em Sov. Phys. JETP}, vol.~1, no.~1, p.~2, 1955.

\bibitem{Watson:1952}
K.~Watson, ``The effect of final state interactions on reaction cross
  sections,'' {\em Phys. Rev. C}, vol.~88, p.~1163, 1952.

\bibitem{Lisa:2005}
M.~Lisa, S.~Pratt, R.~Soltz, and U.~Wiedemann, ``Femtoscopy in relativistic
  heavy ion collisions: Two decades of progress,'' {\em
  Ann.Rev.Nucl.Part.Sci.}, vol.~55, pp.~357--402, 2005.

\bibitem{Shoppa:2000}
T.~D. Shoppa, S.~E. Koonin, and R.~Seki, ``Effect of the source charge on
  charged-boson interferometry,'' {\em Phys. Rev. C}, vol.~61, p.~054902, Apr
  2000.

\bibitem{Tanabashi:2018}
M.~Tanabashi, K.~Hagiwara, K.~Hikasa, K.~Nakamura, Y.~Sumino, F.~Takahashi,
  J.~Tanaka, K.~Agashe, G.~Aielli, C.~Amsler, M.~Antonelli, D.~M. Asner,
  H.~Baer, S.~Banerjee, R.~M. Barnett, T.~Basaglia, C.~W. Bauer, J.~J. Beatty,
  V.~I. Belousov, J.~Beringer, S.~Bethke, A.~Bettini, H.~Bichsel, O.~Biebel,
  K.~M. Black, E.~Blucher, O.~Buchmuller, V.~Burkert, M.~A. Bychkov, R.~N.
  Cahn, M.~Carena, A.~Ceccucci, A.~Cerri, D.~Chakraborty, M.-C. Chen, R.~S.
  Chivukula, G.~Cowan, O.~Dahl, G.~D'Ambrosio, T.~Damour, D.~de~Florian,
  A.~de~Gouv\^ea, T.~DeGrand, P.~de~Jong, G.~Dissertori, B.~A. Dobrescu,
  M.~D'Onofrio, M.~Doser, M.~Drees, H.~K. Dreiner, D.~A. Dwyer, P.~Eerola,
  S.~Eidelman, J.~Ellis, J.~Erler, V.~V. Ezhela, W.~Fetscher, B.~D. Fields,
  R.~Firestone, B.~Foster, A.~Freitas, H.~Gallagher, L.~Garren, H.-J. Gerber,
  G.~Gerbier, T.~Gershon, Y.~Gershtein, T.~Gherghetta, A.~A. Godizov,
  M.~Goodman, C.~Grab, A.~V. Gritsan, C.~Grojean, D.~E. Groom, M.~Gr\"unewald,
  A.~Gurtu, T.~Gutsche, H.~E. Haber, C.~Hanhart, S.~Hashimoto, Y.~Hayato, K.~G.
  Hayes, A.~Hebecker, S.~Heinemeyer, B.~Heltsley, J.~J. Hern\'andez-Rey,
  J.~Hisano, A.~H\"ocker, J.~Holder, A.~Holtkamp, T.~Hyodo, K.~D. Irwin, K.~F.
  Johnson, M.~Kado, M.~Karliner, U.~F. Katz, S.~R. Klein, E.~Klempt, R.~V.
  Kowalewski, F.~Krauss, M.~Kreps, B.~Krusche, Y.~V. Kuyanov, Y.~Kwon,
  O.~Lahav, J.~Laiho, J.~Lesgourgues, A.~Liddle, Z.~Ligeti, C.-J. Lin,
  C.~Lippmann, T.~M. Liss, L.~Littenberg, K.~S. Lugovsky, S.~B. Lugovsky,
  A.~Lusiani, Y.~Makida, F.~Maltoni, T.~Mannel, A.~V. Manohar, W.~J. Marciano,
  A.~D. Martin, A.~Masoni, J.~Matthews, U.-G. Mei\ss{}ner, D.~Milstead, R.~E.
  Mitchell, K.~M\"onig, P.~Molaro, F.~Moortgat, M.~Moskovic, H.~Murayama,
  M.~Narain, P.~Nason, S.~Navas, M.~Neubert, P.~Nevski, Y.~Nir, K.~A. Olive,
  S.~Pagan~Griso, J.~Parsons, C.~Patrignani, J.~A. Peacock, M.~Pennington,
  S.~T. Petcov, V.~A. Petrov, E.~Pianori, A.~Piepke, A.~Pomarol, A.~Quadt,
  J.~Rademacker, G.~Raffelt, B.~N. Ratcliff, P.~Richardson, A.~Ringwald,
  S.~Roesler, S.~Rolli, A.~Romaniouk, L.~J. Rosenberg, J.~L. Rosner, G.~Rybka,
  R.~A. Ryutin, C.~T. Sachrajda, Y.~Sakai, G.~P. Salam, S.~Sarkar, F.~Sauli,
  O.~Schneider, K.~Scholberg, A.~J. Schwartz, D.~Scott, V.~Sharma, S.~R.
  Sharpe, T.~Shutt, M.~Silari, T.~Sj\"ostrand, P.~Skands, T.~Skwarnicki, J.~G.
  Smith, G.~F. Smoot, S.~Spanier, H.~Spieler, C.~Spiering, A.~Stahl, S.~L.
  Stone, T.~Sumiyoshi, M.~J. Syphers, K.~Terashi, J.~Terning, U.~Thoma, R.~S.
  Thorne, L.~Tiator, M.~Titov, N.~P. Tkachenko, N.~A. T\"ornqvist, D.~R. Tovey,
  G.~Valencia, R.~Van~de Water, N.~Varelas, G.~Venanzoni, L.~Verde, M.~G.
  Vincter, P.~Vogel, A.~Vogt, S.~P. Wakely, W.~Walkowiak, C.~W. Walter,
  D.~Wands, D.~R. Ward, M.~O. Wascko, G.~Weiglein, D.~H. Weinberg, E.~J.
  Weinberg, M.~White, L.~R. Wiencke, S.~Willocq, C.~G. Wohl, J.~Womersley,
  C.~L. Woody, R.~L. Workman, W.-M. Yao, G.~P. Zeller, O.~V. Zenin, R.-Y. Zhu,
  S.-L. Zhu, F.~Zimmermann, P.~A. Zyla, J.~Anderson, L.~Fuller, V.~S. Lugovsky,
  and P.~Schaffner, ``Review of particle physics,'' {\em Phys. Rev. D},
  vol.~98, p.~030001, Aug 2018.

\bibitem{Fynbo:2009}
H.~O.~U. Fynbo, R.~\'Alvarez-Rodr\'{\i}guez, A.~S. Jensen, O.~S. Kirsebom,
  D.~V. Fedorov, and E.~Garrido, ``Three-body decays and $r$-matrix analyses,''
  {\em Phys. Rev. C}, vol.~79, p.~054009, May 2009.

\bibitem{Revel:2018}
A.~Revel, F.~M. Marqu\'es, O.~Sorlin, T.~Aumann, C.~Caesar, M.~Holl, V.~Panin,
  M.~Vandebrouck, F.~Wamers, H.~Alvarez-Pol, L.~Atar, V.~Avdeichikov,
  S.~Beceiro-Novo, D.~Bemmerer, J.~Benlliure, C.~A. Bertulani, J.~M. Boillos,
  K.~Boretzky, M.~J.~G. Borge, M.~Caama\~no, E.~Casarejos, W.~N. Catford,
  J.~Cederk\"all, M.~Chartier, L.~Chulkov, D.~Cortina-Gil, E.~Cravo, R.~Crespo,
  U.~Datta~Pramanik, P.~D\'{\i}az~Fern\'andez, I.~Dillmann, Z.~Elekes,
  J.~Enders, O.~Ershova, A.~Estrad\'e, F.~Farinon, L.~M. Fraile, M.~Freer,
  D.~Galaviz, H.~Geissel, R.~Gernh\"auser, P.~Golubev, K.~G\"obel, J.~Hagdahl,
  T.~Heftrich, M.~Heil, M.~Heine, A.~Heinz, A.~Henriques, A.~Ignatov, H.~T.
  Johansson, B.~Jonson, J.~Kahlbow, N.~Kalantar-Nayestanaki, R.~Kanungo,
  A.~Kelic-Heil, A.~Knyazev, T.~Kr\"oll, N.~Kurz, M.~Labiche, C.~Langer,
  T.~Le~Bleis, R.~Lemmon, S.~Lindberg, J.~Machado, J.~Marganiec, A.~Movsesyan,
  E.~Nacher, M.~Najafi, T.~Nilsson, C.~Nociforo, S.~Paschalis, A.~Perea,
  M.~Petri, S.~Pietri, R.~Plag, R.~Reifarth, G.~Ribeiro, C.~Rigollet,
  M.~R\"oder, D.~Rossi, D.~Savran, H.~Scheit, H.~Simon, I.~Syndikus, J.~T.
  Taylor, O.~Tengblad, R.~Thies, Y.~Togano, P.~Velho, V.~Volkov, A.~Wagner,
  H.~Weick, C.~Wheldon, G.~Wilson, J.~S. Winfield, P.~Woods, D.~Yakorev,
  M.~Zhukov, A.~Zilges, and K.~Zuber, ``Strong neutron pairing in {core}+$4n$
  nuclei,'' {\em Phys. Rev. Lett.}, vol.~120, p.~152504, Apr 2018.

\bibitem{Chudoba:2018}
V.~Chudoba, L.~V. Grigorenko, A.~S. Fomichev, A.~A. Bezbakh, I.~A. Egorova,
  S.~N. Ershov, M.~S. Golovkov, A.~V. Gorshkov, V.~A. Gorshkov, G.~Kaminski,
  S.~A. Krupko, I.~Mukha, E.~Y. Nikolskii, Y.~L. Parfenova, S.~I. Sidorchuk,
  P.~G. Sharov, R.~S. Slepnev, L.~Standylo, S.~V. Stepantsov, G.~M.
  Ter-Akopian, R.~Wolski, and M.~V. Zhukov, ``Three-body correlations in direct
  reactions: Example of $^{6}\mathrm{Be}$ populated in the ($p,n$) reaction,''
  {\em Phys. Rev. C}, vol.~98, p.~054612, Nov 2018.

\bibitem{Danilin:1988}
B.~Danilin, M.~Zhukov, A.~Korsheninnikov, V.~Efros, and L.~Chulkov, ``Pauli
  focusing of particles and structure of the ground state of the nucleus $^6$he
  in the $\alpha$+$2n$ model,'' {\em Sov. J. Nucl. Phys.}, vol.~48, p.~766,
  1988.
\newblock [Yad. Fiz. 48, 1208 (1988)].

\bibitem{Zhukov:1993}
M.~V. Zhukov, B.~Danilin, D.~Fedorov, J.~Bang, I.~Thompson, and J.S.Vaagen,
  ``Bound state properties of the borromean halo nuclei: $^6$he and
  $^{11}$li.,'' {\em Phys. Rep.}, vol.~231, pp.~151--199, 1993.

\bibitem{Mei:2012}
P.~Mei and P.~V. Isacker, ``Spatial particle correlations in light nuclei. i
  two-particle systems,'' {\em Annals of Physics}, vol.~327, no.~4, pp.~1162 --
  1181, 2012.

\bibitem{Miernik:2007b}
K.~Miernik, W.~Dominik, Z.~Janas, M.~Pf\"utzner, L.~Grigorenko, C.~R. Bingham,
  H.~Czyrkowski, M.~Cwiok, I.~G. Darby, R.~Dabrowski, T.~Ginter, R.~Grzywacz,
  M.~Karny, A.~Korgul, W.~Kusmierz, S.~N. Liddick, M.~Rajabali, K.~Rykaczewski,
  and A.~Stolz, ``Two-proton correlations in the decay of $^{45}$fe,'' {\em
  Phys. Rev. Lett.}, vol.~99, p.~192501, 2007.

\bibitem{Zhukov:1994}
M.~V. Zhukov, A.~A. Korsheninnikov, and M.~H. Smedberg, ``Simplified
  \ensuremath{\alpha}+4n model for the $^{8}\mathrm{He}$ nucleus,'' {\em Phys.
  Rev. C}, vol.~50, pp.~R1--R4, Jul 1994.

\bibitem{Mei:2012b}
P.~Mei and P.~V. Isacker, ``Spatial particle correlations in light nuclei. ii
  four-particle systems,'' {\em Annals of Physics}, vol.~327, no.~4, pp.~1182
  -- 1201, 2012.

\bibitem{Golubkova:2016}
T.~Golubkova, X.-D. Xu, L.~Grigorenko, I.~Mukha, C.~Scheidenberger, and
  M.~Zhukov, ``Transition from direct to sequential two-proton decay in s-d
  shell nuclei,'' {\em Physics Letters B}, vol.~762, pp.~263--270, 2016.

\bibitem{Azhari:1998}
A.~Azhari, R.~A. Kryger, and M.~Thoennessen, ``Decay of the 12{O} ground
  state,'' {\em Phys. Rev. C}, vol.~58, pp.~2568--2570, Oct 1998.

\bibitem{Brown:2003}
B.~A. Brown and F.~C. Barker, ``Di-proton decay of $^{45}$fe,'' {\em Phys.Rev.
  C}, vol.~67, p.~041304, 2003.

\bibitem{Barker:2003}
F.~C. Barker, ``{$R$}-matrix formulas for three-body decay widths,'' {\em Phys.
  Rev. C}, vol.~68, p.~054602, Nov 2003.

\bibitem{Olsen:2013}
E.~Olsen, M.~Pf\"utzner, N.~Birge, M.~Brown, W.~Nazarewicz, and A.~Perhac,
  ``Landscape of two-proton radioactivity,'' {\em Phys. Rev. Lett.}, vol.~110,
  p.~222501, May 2013.

\bibitem{Mukha:2015}
I.~Mukha, L.~V. Grigorenko, X.~Xu, L.~Acosta, E.~Casarejos, A.~A. Ciemny,
  W.~Dominik, J.~Du\'enas-D\'{\i}az, V.~Dunin, J.~M. Espino, A.~Estrad\'e,
  F.~Farinon, A.~Fomichev, H.~Geissel, T.~A. Golubkova, A.~Gorshkov, Z.~Janas,
  G.~Kami\ifmmode~\acute{n}\else \'{n}\fi{}ski, O.~Kiselev, R.~Kn\"obel,
  S.~Krupko, M.~Kuich, Y.~A. Litvinov, G.~Marquinez-Dur\'an, I.~Martel,
  C.~Mazzocchi, C.~Nociforo, A.~K. Ord\'uz, M.~Pf\"utzner, S.~Pietri,
  M.~Pomorski, A.~Prochazka, S.~Rymzhanova, A.~M. S\'anchez-Ben\'{\i}tez,
  C.~Scheidenberger, P.~Sharov, H.~Simon, B.~Sitar, R.~Slepnev, M.~Stanoiu,
  P.~Strmen, I.~Szarka, M.~Takechi, Y.~K. Tanaka, H.~Weick, M.~Winkler, J.~S.
  Winfield, and M.~V. Zhukov, ``Observation and spectroscopy of new
  proton-unbound isotopes $^{30}\mathrm{Ar}$ and $^{29}\mathrm{Cl}$: An
  interplay of prompt two-proton and sequential decay,'' {\em Phys. Rev.
  Lett.}, vol.~115, p.~202501, Nov 2015.

\bibitem{Xu:2018}
X.-D. Xu, I.~Mukha, L.~V. Grigorenko, C.~Scheidenberger, L.~Acosta,
  E.~Casarejos, V.~Chudoba, A.~A. Ciemny, W.~Dominik, J.~Du\'enas-D\'{\i}az,
  V.~Dunin, J.~M. Espino, A.~Estrad\'e, F.~Farinon, A.~Fomichev, H.~Geissel,
  T.~A. Golubkova, A.~Gorshkov, Z.~Janas, G.~Kami\ifmmode~\acute{n}\else
  \'{n}\fi{}ski, O.~Kiselev, R.~Kn\"obel, S.~Krupko, M.~Kuich, Y.~A. Litvinov,
  G.~Marquinez-Dur\'an, I.~Martel, C.~Mazzocchi, C.~Nociforo, A.~K. Ord\'uz,
  M.~Pf\"utzner, S.~Pietri, M.~Pomorski, A.~Prochazka, S.~Rymzhanova, A.~M.
  S\'anchez-Ben\'{\i}tez, P.~Sharov, H.~Simon, B.~Sitar, R.~Slepnev,
  M.~Stanoiu, P.~Strmen, I.~Szarka, M.~Takechi, Y.~K. Tanaka, H.~Weick,
  M.~Winkler, and J.~S. Winfield, ``Spectroscopy of excited states of unbound
  nuclei $^{30}\mathrm{Ar}$ and $^{29}\mathrm{Cl}$,'' {\em Phys. Rev. C},
  vol.~97, p.~034305, Mar 2018.

\end{thebibliography}

%###############################################################################

\end{document}